\documentclass[12pt,epsbox]{article}
\usepackage{latexsym}
\usepackage{psfig,amssymb,epsf}

\makeatletter
\@addtoreset{equation}{section}
\renewcommand{\theequation}{\thesection.\arabic{equation}}

\newcommand{\bR}{{\bf R}}
\newcommand{\bC}{{\bf C}}

\newcommand{\bZ}{{\bf Z}}

\newcommand{\be}{\begin{equation}} \newcommand{\ee}{\end{equation}}
\newcommand{\bea}{\begin{eqnarray}} \newcommand{\eea}{\end{eqnarray}}
\newcommand{\beq}{\begin{equation}} \newcommand{\eeq}{\end{equation}}

\newcommand{\cA}{{\cal A}}
\newcommand{\cJ}{{\cal J}}

\newcommand{\cP}{{\cal P}}

\newcommand{\del}{{\partial}}

\newcommand{\tr}{{\rm tr}\!}
\newcommand{\trphi}{{\rm tr}_{\phi}}

\newcommand{\nn}{\nonumber \\}

\newcommand{\alpr}{{\alpha^{\prime}}}

\newcommand{\gs}{|0\rangle}

\newcommand{\vthe}{\vartheta}
\newcommand{\cthe}[2]{\vartheta\left[
                        \begin{array}{c}#1\\
                                        #2
                        \end{array}\right]}

\font\zfont = cmss10 
\newcommand{\ZZ}{\hbox{\zfont Z\kern-.4emZ}}

\ifx\TwoupWrites\UnDeFiNeD\else\target{\magstepminus1}{11.3in}{8.27in}
\source{\magstep0}{7.5in}{11.69in}\fi
\newfont{\fourteencp}{cmcsc10 scaled\magstep2}
\newfont{\titlefont}{cmbx10 scaled\magstep3}
\newfont{\authorfont}{cmcsc10 scaled\magstep1}
\newfont{\fourteenmib}{cmmib10 scaled\magstep2}
\skewchar\fourteenmib='177
\newfont{\elevenmib}{cmmib10 scaled\magstephalf}
\skewchar\elevenmib='177
\makeatletter
\newcommand\nonsequentialeqnum{
\@addtoreset{equation}{section}
\def\theequation{\arabic{section}.\arabic{equation}}}
\newif\ifp@bblock \p@bblocktrue
\newcommand\nopubblock{\p@bblockfalse}
\newcommand\topspace{\hrule height 0pt depth 0pt \vskip}
\newcommand\p@bblock{\begingroup \tabskip=\hsize minus \hsize
\baselineskip=1.5\ht\strutbox \topspace-2\baselineskip
\halign to\hsize{\strut ##\hfil\tabskip=0pt\crcr
\the\Pubnum\crcr\the\date\crcr}\endgroup}

\renewcommand\titlepage{\ifx\TwoupWrites\UnDeFiNeD\null
\vspace{-1.7cm}\fi
\vskip0.6cm
\ifp@bblock\p@bblock \else\hrule height 0pt \relax \fi}
\makeatother
\newtoks\date
\newtoks\Pubnum
\newtoks\pubnum
\newcommand{\frontpageskip}{\vspace{12pt plus .5fil minus 2pt}}
\renewcommand{\title}[1]{\frontpageskip
\begin{center}{\titlefont #1}\end{center}\par}
\renewcommand{\author}[1]{\frontpageskip\par\begin{center}
{\authorfont #1}\end{center}
\nobreak
}

\renewcommand{\thanks}[1]{\footnote{#1}}
\renewcommand{\abstract}{\par\frontpageskip\centerline{
\fourteencp Abstract}
\vspace{8pt plus 3pt minus 3pt}}

\begin{document}

\begin{titlepage}
\hfill
\vbox{
    \halign{#\hfil         \cr
           TAUP-2779-04 \cr
           IU-MSTP-67 \cr
           hep-th/0503043  \cr
           } 
      }  
\vspace*{20mm}
\begin{center}
{\Large {\bf $N=2$ Strings on Orbifolds}\\} 
\vspace*{15mm}
{\sc Dan Gl\"uck}{$^a\!$}
\footnote{e-mail: {\tt gluckdan@post.tau.ac.il}}
{\sc Yaron Oz}{$^a\!$} 
\footnote{e-mail: {\tt yaronoz@post.tau.ac.il}}
and {\sc Tadakatsu Sakai}{$^b\!$}
\footnote{e-mail: {\tt tsakai@mx.ibaraki.ac.jp}}

\vspace*{1cm} 
{\it {$^{a}$ Raymond and Beverly Sackler Faculty of Exact Sciences\\
School of Physics and Astronomy\\
Tel-Aviv University , Ramat-Aviv 69978, Israel}}\\ 
\vspace*{.5cm} 
{\it {$^{b}$
Department of Mathematical Sciences, Faculty of Sciences,\\
Ibaraki University, Bunkyo 2-1-1, Mito, 310-8512 Japan
}}\\

\end{center}

\begin{abstract}
We study closed  
$N=2$
strings  on orbifolds of the form $T^4/\bZ_2$ and $\bC^2/\bZ_2$.
We compute the torus partition function and prove its modular invariance.
We analyze the BRST cohomology of the theory, construct
the vertex operators, and 
compute three and four point amplitudes of twisted and untwisted 
states.
We introduce a  background
of D-branes, and 
compute twist states correlators.

\end{abstract}
\vskip .7cm

March 2005

\end{titlepage}

\setcounter{footnote}{0}

\newpage

\section{Introduction}

Closed $N=2$ strings \cite{ademollo;76, Gates:1988tn} possess local $N=2$
supersymmetry on the string worldsheet.
Critical $N=2$ strings have a four-dimensional target space
\cite{Ooguri:1991fp}.
The supersymmetric structure implies that the target space
has a complex structure. Therefor it must be of signature $(4,0)$
or $(2,2)$.
In $(4,0)$ signature there are no propagating degrees of freedom
in the $N=2$ string spectrum.
In $(2,2)$ signature there are only massless scalars  in the
spectrum, and
the infinite tower of massive excitations of the string is absent.

Closed $N=2$ strings have been divided to two T-dual families denoted
by $\alpha$ and $\beta$ \cite{Cheung:2002yw}.
$\beta$ strings are the ones studied in \cite{Ooguri:1991fp}, where it was
shown that  
they have self-dual (curvature) four-manifolds
as their target space.
The  $\alpha$ strings effective action has been constructed in
\cite{Gluck:2003wg}, where it was shown that the target space 
dynamics is that of self-dual (curvature) four-manifolds with torsion
\cite{Hull:1996zt}.
D-branes in the $N=2$ strings have been analyzed in
\cite{Cheung:2002yw,Gluck:2003pa}.
$N=2$ strings have global $N=4$ worldsheet supersymmetry \cite{Ohta:1989pr}.
They have been shown to be equivalent to topological $N=4$ strings in
\cite{Berkovits:1994vy}.
They have been suggested as describing the topological
sector of six-dimensional little string theories with sixteen 
supercharges 
\cite{Aharony:2003vk}.

Despite the simplicity of $N=2$ strings, very little is known 
when the target space is curved. Curved backgrounds are
of particular importance, when 
looking for open-closed string duality structure in the $N=2$ strings.
In this paper we will take one step in this direction, and
study $N=2$ strings
on orbifolds of the form $T^4/\bZ_2$ and $\bC^2/\bZ_2$.
First, we will compute the torus partition function and prove its modular invariance.
We will then analyze the BRST cohomology of the theory in different pictures.
We will see
that the different pictures are not isomorphic. 
We will find a massless scalar and global
degrees of freedom
in the untwisted sector, and one twisted
state in
the twisted sector.
We will then construct
the vertex operators, 
compute three and four point amplitudes of twisted and untwisted 
states, and introduce the low-energy effective action.
We will then add a background of D-branes,
compute twist states correlators in their background, consider the effective world volume theory
and compare it with the unorbifolded case.
Finally, we will discuss the results.

The paper is organized as follows.
In section 2 we will consider the torus partition function
and show that it is  modular invariant.
We analyze the spectrum and construct
the vertex operators.
In section 3 we compute three and four point amplitudes of twisted and untwisted 
states, and interpret the results.
In section 4 we will consider closed strings in D-branes background and
compute twist states correlators. We discuss the effective world volume theory 
and compare it with the unorbifolded case.

\section{Partition function and Spectrum of the $N=2$ Orbifold}

In this section we
introduce the torus partition function and show that it 
is modular invariant.
We then introduce
the spectrum of the $N=2$ string on the $\bZ_2$ orbifold.
We will discuss shortly the spectrum of the untwisted sector, which consists of 
a massless scalar and global degrees of freedom (graviton, dilaton and a 2-form).
We will also perform a BRST analysis in the twisted sector and show the spectrum consists of a single state.
We will then construct the corresponding vertex operators.

For an earlier work on the partition function of the $N=2$ string, see \cite{bl:1996}.
For an earlier work on BRST analysis in $N=2$ string orbifolds,
see \cite{Mathur:1986cz,klp}.
The interested reader may find detailed derivations of the results in appendices B and C.

\subsection{Partition Function}

The partition function is
\begin{equation}
d^2\tau d\theta\, \trphi \left(\cP \, q^{L_0}\bar{q}^{\bar{L}_0} e^{-2\pi i\theta (J_0+\bar{J}_0)}\right)
\end{equation}
where $\phi$ and $2\pi\theta$ are the periodicities over the $\alpha$ and $\beta$ cycles of the torus, respectively,
of any oscillator with a $U(1)$-charge $1$.

For the $\bZ_2$ orbifold, the projection operator is given by
$\cP=(1+{\bf r})/2$ where ${\bf r}$ is the $\bZ_2$ reflection.

\subsubsection{Oscillators}

For the computation of the $N=2$ string partition function we 
consider two complex scalars, two holomorphic and two antiholomorphic
complex fermions, two sets of holomorphic and
antiholomorphic anticommuting ghosts and two sets of holomorphic and
antiholomorphic commuting superghosts.

The sum over bosonic oscillators is similar to that in the bosonic string, and
can be found in section 2.1 of appendix B.

The fermions give
\begin{eqnarray}
\trphi \left(q^{L_0}\bar{q}^{\bar{L}_0} e^{-2\pi i\theta (J_0+\bar{J}_0)}\right) 
&=& {1\over\left|\eta(\tau)\right|^4} \left|\cthe{\phi}{-\theta}(0,\tau)\right|^4 \ , \nn
\nn
\trphi \left({\bf r}\cdot q^{L_0}\bar{q}^{\bar{L}_0} e^{-2\pi i\theta (J_0+\bar{J}_0)}\right)
&=& {1\over\left|\eta(\tau)\right|^4} \left|\cthe{\phi}{-\theta\!+\!1/2}(0,\tau)\right|^4 . \nn
\end{eqnarray}

In the twisted sector 
\begin{eqnarray}
\trphi \left(q^{L_0}\bar{q}^{\bar{L}_0} e^{-2\pi i\theta (J_0+\bar{J}_0)}\right) 
&=& {1\over\left|\eta(\tau)\right|^4}
\left|\cthe{\phi\!+\!1/2}{-\theta}(0,\tau)\right|^4 \ , \nn
\nn
\trphi \left( {\bf r} \cdot q^{L_0}\bar{q}^{\bar{L}_0} e^{-2\pi i\theta (J_0+\bar{J}_0)}\right)  
&=& {1\over\left|\eta(\tau)\right|^4}
\left|\cthe{\phi\!+\!1/2}{-\theta\!+\!1/2}(0,\tau)\right|^4 \ . \nn
\end{eqnarray}

For the ghost and superghost sectors one obtains, as in the unorbifolded $N=2$ string,
\begin{eqnarray}
\trphi \left( (-1)^{F_{\rm gh}}b_0 c_0\bar{b}_0\bar{c}_0\,\hat{b}_0\hat{c}_0\bar{\hat{b}}_0\bar{\hat{c}}_0\,
q^{L_0}\bar{q}^{\bar{L}_0} e^{-2\pi i\theta (J_0+\bar{J}_0)}\right)
&\!=\!&
\left|\eta(\tau)\right|^{12} \, \left|\cthe{\phi}{-\theta}(0,\tau)\right|^{-4} \ .
\end{eqnarray}

\subsection{Modular Invariance}

Collecting these results together, one obtains
\begin{equation}
Z_{\rm orbifold}={1\over 2}\int d^2\tau \,d\theta d\phi\left[Z(\tau,\bar{\tau})+I\right] \ ,
\label{partfunc}
\end{equation}
where $Z(\tau,\bar{\tau})$ is the partition function of the unorbifolded
$N=2$ string,
\begin{equation}
Z(\tau,\bar{\tau}) = {V\over (4\pi^2\alpha^{\prime}\tau_2)^2} \ ,
\end{equation}
in the non-compact case, and
\begin{equation}
Z(\tau,\bar{\tau})= \sum_{p\in\Gamma} q^{{\alpha^{\prime}\over 4}p_L^2} \bar{q}^{{\alpha^{\prime}\over 4}p_R^2} \ ,
\end{equation}
in  the compact case, where $\Gamma$ is the lattice of allowed momenta.

In  the non-compact case:
\begin{equation}
I=
{\left|\eta(\tau)\right|^{12}\over \left|\cthe{\phi}{-\theta}(0,\tau)\right|^4}
\left(
{\left|\cthe{\phi}{-\theta\!+\!1/2}(0,\tau)\right|^4\over \left|\vthe_{10}(0,\tau)\right|^4}
+
{\left|\cthe{\phi\!+\!1/2}{-\theta}(0,\tau)\right|^4\over
 \left|\vthe_{01}(0,\tau)\right|^4}
+{\left|\cthe{\phi\!+\!1/2}{-\theta\!+\!1/2}(0,\tau)\right|^4\over
 \left|\vthe_{00}(0,\tau)\right|^4}
\right) \ .
\end{equation}
In the compact case $I$ should be multiplied by an overall factor of $2^4$.

$Z(\tau,\bar{\tau})$ is known to be modular-invariant after integration by $d\tau d\bar{\tau}$, both 
for the non-compact case and for the compact case (for the latter this can be shown 
by Poisson resummation, for an appropriate lattice $\Gamma$).

We will now show that $I$ is modular invariant after integration by $d\tau d\bar{\tau}d\theta d\phi$.

Under $\tau\rightarrow\tau+1$, The second and the third
terms interchange while the first term remains invariant. This can be shown by taking $\theta\rightarrow\theta-\phi-1/2$ and using
\begin{eqnarray}
&\cthe{a}{b}(0,\tau+1)=e^{-\pi ia(a+1)}\,\cthe{a}{a+b+1/2}(0,\tau) \ ,&
\nn\nn
&\cthe{a+1}{b}(0,\tau) = \cthe{a}{b}(0,\tau) \,\, \ , \,\, \cthe{a}{b+1}(0,\tau) = e^{2\pi ia}\cthe{a}{b}(0,\tau) \ ,&
\label{ctheinv}
\end{eqnarray}
Thus $Z_{\rm orbifold}$ is invariant under $\tau\rightarrow\tau+1$.

Also, one can see that under $\tau\rightarrow-1/\tau$ the first and second terms 
interchange while the third term 
remains invariant, by taking $\theta \rightarrow \phi$ and $\phi \rightarrow -\theta$ and using (\ref{ctheinv}) and 
\begin{equation}
\cthe{a}{b}(0,-1/\tau) = (-i\tau)^{1/2}e^{2\pi iab}\,\cthe{b}{-a}(0,\tau) \ . 
\end{equation}
Thus $Z_{\rm orbifold}$ is invariant under $\tau\rightarrow-1/\tau$.

\subsection{Spectrum of the $N=2$ Orbifold}

We begin by discussing different pictures in the $N=2$ string.
As in the $N=1$ string, a picture is defined by the superghost sector ground state \cite{Friedan:1985ge}.
In the $N=2$ string there are two sets of superghosts, and the picture is defined by a pair of numbers rather than 
a single number.
The pictures that will be most relevant for us are $(-1,-1)$, $(0,0)$ (for NS boundary conditions) and $(-1/2,-1/2)$
(for R boundary conditions), defined in appendix C.

The picture changing operator relates physically equivalent states of different pictures.
As has been discussed in \cite{JL:1997}, the picture changing operator is an isomorphism for non-zero momentum states,
but not for zero momentum states. Thus, different pictures may have different states with vanishing momentum.

Another isomorphism between physically equivalent 
states of different pictures is the spectral flow, which is discussed in section 3 of appendix C. 

Because of these isomorphisms, for states with non-vanishing momentum it is sufficient to study 
the BRST cohomology in the $(-1,-1)$ picture, which consists only of a massless scalar. 
Thus the spectrum of the unorbifolded $N=2$ string consists of a massless scalar, and zero momentum states 
which turn out to describe dilaton, metric and B-field global degrees of freedom 
\cite{Bienkowska:1991zs} (see section 1 of appendix C).
The untwisted sector of the $N=2$ orbifold at hand consists of $\bZ_2$ invariant combinations of these states.

\vskip 0.3cm
\noindent
{\bf BRST Analysis of the Twisted Sector}

In the twisted sector there are no momentum degrees of freedom, and the BRST cohomology is different
in different pictures:
\begin{itemize}
\item
In the $(-1,-1)$ picture
the normal ordering constant of $L_0$ is $+1/2$,
and The BRST cohomology is empty.

\item
In the $(-1/2,-1/2)$ picture $L_0$ has zero normal ordering constant, 
and the ground state is the only physical state.
The corresponding vertex operator is given by
\begin{equation}
c \, e^{-\phi^+/2} \, e^{-\phi^-/2} \, (z) \, 
\bar{c} \, e^{-{\bar \phi}^+/2} \, e^{-{\bar \phi}^-/2} \, (\bar{z})\cdot \, 
\sigma^1(z,\bar{z})\,\sigma^2(z,\bar{z}) \ .
\end{equation}
where $\sigma^i$ is the twist field creating the cut in $X^i$ and $\bar{X}^i$ in the $z$-plane \cite{Dixon:1987},
and $\beta^{\mp}\gamma^{\pm} = \partial \phi^{\pm}$ \cite{JL:1997}.

\item
In the $(0,0)$ picture
the normal ordering constants of $L_0$ is
$-1/2$, and we get two states in the BRST cohomology, whose matter parts read 
\begin{eqnarray}
G^{-~M}_{-1/2} \psi^1_0 \psi^2_0 |\sigma\rangle_{(0,0)} \ &,&
G^{+~M}_{-1/2} |\sigma\rangle_{(0,0)} \ .
\end{eqnarray}
These states are physically equivalent to the twisted ground state in the $(-1/2,-1/2)$ picture, 
through the use of the picture changing operator and spectral flow (see section 3 of appendix C for details).

\end{itemize}

\section{Scattering Amplitudes on $\bC^2/\bZ_2$ and $T^4/\bZ_2$}

In this section we will compute scattering amplitudes of untwisted and twisted states in
$\bC^2/\bZ_2$ and $T^4/\bZ_2$ backgrounds.
We work with $\alpha^\prime=2$ in this section.

\subsection{Four Twisted States Amplitude}

A twisted state is located in target space at a fixed point
of the orbifold group.
For the $T^4/\bZ_2$ orbifold with radius $R$, they are at 
$\pi R\epsilon$, where $\epsilon$ is a four-dimensional vector (in real coordinates)
such that $\epsilon^\mu\in\{0,1\}$. 
For twisted states located at the fixed points $2\pi f_{\epsilon_i}$, $i=1,...4$, 
($f_{\epsilon_i}= {1\over 2} R\epsilon_i$) the four twisted states scattering amplitude is
\begin{equation}
{\cal A}_{tttt} = \int dx d\bar{x} \, \langle c\bar{c}V^{(-{1\over 2},-{1\over 2})}_{\epsilon_4} (z_\infty) \,
c\bar{c}V^{(-{1\over 2},-{1\over 2})}_{\epsilon_3}(1) \, V^{(-{1\over 2},-{1\over 2})}_{\epsilon_2}(x,\bar{x}) \,
c\bar{c}V^{(-{1\over 2},-{1\over 2})}_{\epsilon_1}(0) \rangle_{S^2} \ ,
\label{A4}
\end{equation}
where 
\begin{equation}
V^{(-{1\over 2},-{1\over 2})}_{\epsilon_i} (z,\bar{z}) = 
e^{-\phi^+/2} \, e^{-\phi^-/2} \, (z) \cdot e^{-\bar{\phi}^+/2} \, e^{-\bar{\phi}^-/2} \, (\bar{z})
\cdot \, \sigma^1_{\epsilon_i}(z,\bar{z})\,\sigma^2_{\epsilon_i}(z,\bar{z}) \ .
\end{equation}
$\sigma^j_{\epsilon_i}$ is a twist field for $X^j$, located at $2\pi f_{\epsilon_i}$.

The matter part of the integrand in (\ref{A4}) is
\begin{equation}
\langle \sigma^1_{\epsilon_4}\sigma^2_{\epsilon_4}(z_\infty) \, \sigma^1_{\epsilon_3}\sigma^2_{\epsilon_3}(1) \, 
\sigma^1_{\epsilon_2}\sigma^2_{\epsilon_2}(x,\bar{x}) \, \sigma^1_{\epsilon_1}\sigma^2_{\epsilon_1}(0)\rangle_{S^2} \ .
\end{equation}
Define
\begin{equation}
Z(x,\bar{x}) \equiv \lim_{z_\infty\rightarrow\infty} |z_\infty| \, 
\langle \sigma^1_{\epsilon_4}\sigma^2_{\epsilon_4}(z_\infty) \, \sigma^1_{\epsilon_3}\sigma^2_{\epsilon_3}(1) \, 
\sigma^1_{\epsilon_2}\sigma^2_{\epsilon_2}(x,\bar{x}) \, \sigma^1_{\epsilon_1}\sigma^2_{\epsilon_1}(0)\rangle_{S^2} \ .
\end{equation}
$Z$ can be recast as $Z=Z_1Z_2$, where $Z_i$ denotes a
four-point function of twist fields on $T^2/\bZ_2$ (or $\bC/\bZ_2$).

\subsubsection{$T^4/\bZ_2$ Orbifold}

Define $T^4$ as $\bR^4/2\pi\Lambda$ with $2\pi\Lambda$ being
a four-dimensional lattice.
When all the radii of $T^4$ are equal to $R$, 
$\Lambda=\{(m_1R,m_2R,m_3R,m_4R)|m_i\in\bZ\}$.
For every fixed point $2\pi f_\epsilon$, $2 f_\epsilon$ is an element of $\Lambda$.
Global monodromy conditions  (changes of $X$ along closed loops) imply that (\ref{A4}) is non-zero only if
$\sum{f_{\epsilon_i}}$ belongs to $\Lambda$ 
as well \cite{Dixon:1987}.

On $T^4/\bZ_2$ one gets \cite{Dixon:1987}
\begin{eqnarray}
Z_i(x,\bar{x}) &=& {{|x(1-x)|}^{-1/2}\over {\tau_2(x,\bar{x})|F(x)|}^2}\, V_{\Lambda_i}
\sum_{v_1\in{\Lambda_i}_c,v_2\in{\Lambda_i}_c'}e^{-{\pi\over 4\tau_2}[v_2\bar{v}_2-\tau_1(v_1\bar{v}_2+\bar{v}_1v_2)+
|\tau|^2v_1\bar{v}_1]} \ ,
\label{zi_comp}
\end{eqnarray}
where $\tau(x) \equiv iF(1-x)/F(x)$, and $F(x)$ is the hypergeometric function
\begin{equation}
F(x)\equiv F(1/2,1/2,1;x)={1\over\pi}\int_0^1 dy \, y^{-1/2} (1-y)^{-1/2} (1-xy)^{-1/2} \ .
\end{equation}
We denoted 
$\Lambda_c\equiv (1-\theta)(f_{\epsilon_2}-f_{\epsilon_1}+\Lambda)$ and 
$\Lambda_{c'}\equiv (1-\theta)(f_{\epsilon_2}-f_{\epsilon_3}+\Lambda)$, where $\theta$ is the $\bZ_2$ rotation.
$V_{\Lambda}$ is the unit cell volume in $\Lambda$.
The index $i$ in ${\Lambda_i}_c$, ${\Lambda_i}_{c'}$, $f_{\epsilon_j}^i$, $V_{\Lambda_i}$ stands for a projection onto the $i$th dimension.

The holomorphic ghost part of the four point function reads
\begin{eqnarray}
&\langle e^{-{{\phi}^+}/2}(z_\infty) e^{-{{\phi}^+}/2}(1) e^{-{{\phi}^+}/2}(x) e^{-{{\phi}^+}/2}(0) \rangle_{S^2}
\langle e^{-{{\phi}^-}/2}(z_\infty) e^{-{{\phi}^-}/2}(1) e^{-{{\phi}^-}/2}(x) e^{-{{\phi}^-}/2}(0) \rangle_{S^2} &\nn
&\cdot\langle c(z_\infty) c(1) c(0) \rangle_{S^2} = {z_\infty}^{1/2} x^{-1/2}(1-x)^{-1/2}& \ ,
\end{eqnarray}
with a similar expression for the anti-holomorphic part.
The total four point function is
\begin{equation}
{\cal A}_{tttt}=\int dx d\bar{x} \, {{|x(1-x)|}^{-2}\over \tau_2^2(x,\bar{x})|F(x)|^4} 
V_{\Lambda} \sum_{v_1\in\Lambda_c,v_2\in\Lambda_{c'}}
e^{-{\pi\over 4\tau_2}[v_2\bar{v}_2-\tau_1(v_1\bar{v}_2+\bar{v}_1v_2)+
|\tau|^2v_1\bar{v}_1]}\ ,
\label{4ptfun_comp}
\end{equation}
where $v_{1,2}$ are two complex dimensional vectors.

It is further shown in section 1 of appendix D that ${\cal A}_{tttt}$ is finite and invariant under crossing symmetry.

Let us discuss this result.
The sum over $v_1$ corresponds to a sum
over global monodromy conditions
for a contour surrounding $0$ and $x$. In the 
$s$-channel ($x\sim 0$), it represents a sum over intermediate winding states
(that wind around $f_{\epsilon_1}$ and 
$f_{\epsilon_2}$).
The sum over $v_2$ corresponds to a
sum over global monodromy conditions for a contour surrounding $x$ and $1$.
We can use 
Poisson resummation and change variables to $p$, 
representing the momentum of the intermediate states
in the $s$-channel, as in \cite{Dixon:1987}. 
Then (\ref{4ptfun_comp}) takes the form
\begin{equation}
{\cal A}_{tttt}= \int dx d\bar{x} {{|x(1-x)|}^{-2}\over {|F(x)|}^4} 
\sum_{p\in\Lambda^*,v\in\Lambda_c} \exp\left(-2\pi i (f_{\epsilon_2}-f_{\epsilon_3})\cdot p \right) 
w^{(p+v/2)^2/2}\bar{w}^{(p-v/2)^2/2} \ ,
\label{Poisson}
\end{equation}
where $w(x)\equiv e^{i\pi\tau(x)}$ and $\Lambda^*$ is the dual lattice of $\Lambda$.
For instance, when all the $T^4$ radii 
are equal, then $\Lambda^*=\{(m_1/R,m_2/R,m_3/R,m_4/R)|m_i\in\bZ\}$.
Note that the $e^{-2\pi i (f_{\epsilon_2}-f_{\epsilon_3})\cdot p}$ phase is always $\pm 1$.
In every term of the sum (\ref{Poisson}), $p$ represents the 
momentum of the intermediate untwisted state in the $s$-channel
and $v/R$ is its winding number. Thus,
the left and right momenta are $p_{L,R}=p\pm v/2$. 

${\cal A}_{tttt}$ seems to have poles only at 
$p_L^2 = p_R^2 = 0$ (see section 2 of appendix D).
Performing the integral in (\ref{Poisson}) over a disk of radius $a$ around $x=0$ gives
\begin{eqnarray}
2\pi \, {(a/16)^{p_L^2}\over p_L^2} \, \delta_{p_L^2-p_R^2} + f(a,p_L^2,p_R^2)
\label{uttutt}
\end{eqnarray}
where $f$ has no poles in $p_L^2$, $p_R^2$.

This suggests that in the effective field theory description
only poles of intermediate physical states contribute to the scattering amplitude.
However, there may be an additional contact term contribution to the amplitude. 

\subsubsection{$\bC^2/\bZ_2$}

In the non-compact case the global monodromy conditions are trivial, and there is only one fixed point
at the origin. 
We have
\begin{equation}
Z_i(x,\bar{x}) = {2{|x(1-x)|}^{-1/2}\over F(x)\bar{F}(1-\bar{x})+F(1-x)\bar{F}(\bar{x})} = 
{{|x(1-x)|}^{-1/2}\over \tau_2(x,\bar{x}){|F(x)|}^2} \ ,
\label{zi_nc}
\end{equation}
up to a normalization constant of $(2\pi)^{-2}$ which is justified below. The total four point function reads
\begin{equation}
{\cal A}_{tttt} = \int dx d\bar{x} Z_1(x,\bar{x}) Z_2(x,\bar{x}) {|x(1-x)|}^{-1} = 
(2\pi)^{-4} \int dx d\bar{x} {{|x(1-x)|}^{-2}\over \tau_2^2(x,\bar{x}){|F(x)|}^4} \ ,
\label{4ptfun_nc}
\end{equation}
which
is identical to (\ref{4ptfun_comp}) for trivial lattices $\Lambda_c$, $\Lambda_{c'}$, and is finite.

Rewriting $1/\tau_2$ as
\begin{equation}
{1\over \tau_2(x,\bar{x})} = 
\int dp_a dp_b \, e^{-\pi ({p_a}^2 + {p_b}^2) \tau_2} = 
\int d^2p \, w^{p^2/2}\bar{w}^{p^2/2} \ ,
\end{equation}
we get 
\begin{equation}
{\cal A}_{tttt}= \int dx d\bar{x} \, {{|x(1-x)|}^{-2}\over {|F(x)|}^4} \int {d^4p\over (2\pi)^4} \, w^{p^2/2}\bar{w}^{p^2/2} \ .
\label{ncPois}
\end{equation}
As in the $T^4/\bZ_2$ case, this can be interpreted as a
sum over intermediate states with poles at $p^2=0$. 
We normalized the amplitude by demanding that for $p=0$ the leading term in (\ref{ncPois}) at $x\rightarrow 0$ is 
$1\cdot |x|^{-2}$ \cite{Dixon:1987}.

\subsection{Twisted-Twisted-Untwisted States Amplitude}

The scattering amplitude of two twisted states and one untwisted state,
${\cal A}_{utt}$, can be derived using the factorization property
of ${\cal A}_{tttt}$.
In fact, as seen in the previous section,
the residue of the pole in (\ref{uttutt}) is equal to 
${\cal A}_{utt}^2$. Hence
\begin{equation}
{\cal A}_{utt}=\sqrt{2\pi} \ .
\label{utt}
\end{equation}

Equation (\ref{utt}) can also be obtained using the state-operator correspondence. 
When the twisted states are 
located on the same fixed point, we have 
\beq
\langle \sigma | {1\over 2} (e^{i (p_L\cdot X^L + p_R\cdot X^R)} + 
e^{-i (p_L\cdot X^L + p_R\cdot X^R)}) |\sigma\rangle 
= 1 \ .
\label{op3}
\eeq
Here $p\cdot X$ means $p_i\bar{X}^i + \bar{p}_i X^i$.
Note that there is no momentum conservation, since there is no target-space translation invariance. 
This is seen in (\ref{op3}) by the absence of the center-of-mass mode $x^\mu$, which would give a 
delta function in $p$ in the unorbifolded string.

The discussion generalizes to a non-compact target space $\bC^2$, by the use of the representation 
(\ref{ncPois}).
We conclude that on-shell, the 3-point function of two twisted states and one untwisted state is 
independent of momentum, winding number, or the twisted states location, and is equal to one.

In order to construct the effective action involving the twisted state,
we need information of the off-shell $ttu$ vertex.
Here we assume that it is constant as well. 
The reason for this assumption will be clearer later on.

\subsection{Two Twisted and Two Untwisted States Amplitude}

We will consider the
two twisted - two untwisted states amplitude when
the two twisted states are located at the same fixed point.
The amplitude is computed as a correlation 
function of the two untwisted vertex operators, using their Green function
in the presence of twist-fields.
This method can be used when the total winding number of the two
untwisted states is zero.

In the $(0,0)$ picture, the matter part of the untwisted state vertex operator is
\beq
{1\over 2}V^{(0,0)}(k_{L,R};z,\bar{z})+{1\over 2}V^{(0,0)}(-k_{L,R};z,\bar{z}) \ ,
\eeq
where $V^{(0,0)}(k_{L,R};z,\bar{z})$ is given by \cite{Gluck:2003wg}
\begin{equation}
V^{(0,0)}(k_{L,R};z,\bar{z}) =
\left( -k_L \cJ\partial X+{1\over 4} \, \left(k_L J_+\psi\right) \, \left(k_L J_-\psi\right)\right) \, 
e^{ik_L\cdot X^L}\cdot (h.c) \ .
\end{equation}
$\cJ$ is the complex structure and $J_\pm \equiv \eta\pm i\cJ$.

The two twisted - two untwisted states
amplitude, for untwisted states with momenta $\pm k_{L,R}$ and $\pm p_{L,R}$, is
\begin{eqnarray}
{\cal A}_{tt,p,k}&=&\int dzd\bar{z} \, \langle \sigma_1(z_\infty)\sigma_2(z_\infty) \,
{1\over 2}\left(e^{ik_{L,R}\cdot X^{L,R} (1)} + e^{-ik_{L,R}\cdot X^{L,R} (1)}\right) \, \nn
&& {1\over 2}\left(V^{(0,0)}(p_{L,R};z,\bar{z})+V^{(0,0)}(-p_{L,R};z,\bar{z})\right) \, \sigma_1(0)\sigma_2(0)\rangle \nn
&&\cdot|\langle e^{-\phi^+(z_\infty)/2}e^{-\phi^+(1)}e^{-\phi^+(0)/2}\rangle \,
\langle e^{-\phi^-(z_\infty)/2}e^{-\phi^-(1)}e^{-\phi^-(0)/2}\rangle \, \langle c(z_\infty)c(1)c(0)\rangle|^2  \ ,
\nn
\end{eqnarray}
where $k_{L,R}\cdot X^{L,R} \equiv k_L \cdot X^L + k_R \cdot X^R$.

By using the bosonic fields Green function in the presence of two twist fields we get (see appendix E for details)
\begin{equation}
{\cal A}_{tt,p,k}= -2\left(p_L\cJ k_L\right)\left(p_R\cJ k_R\right) 
\bigg[I\left(2k_L\cdot p_L,2k_R\cdot p_R\right) + 
I\left(-2k_L\cdot p_L,-2k_R\cdot p_R\right)\bigg] \ ,
\label{AII}
\end{equation}
where
\begin{eqnarray}
I(a_L, a_R) &\equiv& \int dzd\bar{z} \, |z(1-z)^2|^{-1} 
\left({1-z^{1/2}\over 1+z^{1/2}}\right)^{a_L} \left({1-\bar{z}^{1/2}\over 1+\bar{z}^{1/2}}\right)^{a_R} \nn
&=& \int_{{\rm Im} (v)\ge 0} dvd\bar{v} \, (1-v)^{a_L-1} v^{-a_L-1} (1-\bar{v})^{a_R-1} \bar{v}^{-a_R-1} \ .
\end{eqnarray}
When $a_L-a_R$ is an integer we may write
\begin{eqnarray}
I(a_L, a_R) + I(-a_L, -a_R) &=& 
\int_{\bC} dvd\bar{v} \, (1-v)^{a_L-1} v^{-a_L-1} (1-\bar{v})^{a_R-1} \bar{v}^{-a_R-1} \nn
&=& \pi {\Gamma(-a_L)\Gamma(a_L)\Gamma(1) \over \Gamma(0) \Gamma(1-a_R) \Gamma(1+a_R)} = 0 \ .
\label{IaLaR}
\end{eqnarray}

Since $k_L \cdot p_L - k_R \cdot p_R$ must be an integer for the theory to be modular invariant 
(as in any string theory with a toroidal compactification), we can  use (\ref{IaLaR}) to calculate the 
scattering amplitude (\ref{AII}). We thus get
\begin{eqnarray}
{\cal A}_{tt,p,k} &=& 0 \ .
\end{eqnarray}
All the above is
easily generalized to the non-compact background $\bC^2/\bZ_2$ by setting $p_L=p_R$, 
$k_L=k_R$.

Study of the function $I(a_L,a_R)$ (see section 1 of appendix E) suggests the following effective 
field theory description.
Only two diagrams contribute to the amplitude,
both are of the form $ttu$ - $uuu$, but in one of them the intermediate 
untwisted state has momentum $\pm(p_{L,R}+k_{L,R})$, and in the other it has momentum $\pm(p_{L,R}-k_{L,R})$. 
Recall that the momentum of an untwisted state on the orbifold is defined only up to a sign.
Only the physical poles (with $(p_{L,R}\pm k_{L,R})^2=0$) appear, 
and there is no two twisted - two untwisted states contact term.
The contributions of the two diagrams are 
$\left[\left(p\cJ k\right)\left(p\cJ k\right)-\left((p\pm k)^2\right)^2\right] / (p\pm k)^2$.
In fact,  the
$u(p_1)u(p_2)u(p_3)$ vertex with $p_1,p_2$ on-shell and $p_3$ off-shell
becomes $(p_1\cJ p_2)^2-(p_3^2)^2$ \cite{Ooguri:1991fp}.
Also we have assumed that the off-shell $ttu$ vertex is constant.
By noting $(p-k)^2 = -(p+k)^2$,
one finds that
the two diagrams cancel each other.
For the similar reasons, we conjecture that the scattering amplitude of two
untwisted states with two twisted states vanishes always, including
the cases where the total untwisted states winding number is non-zero,
and where the twisted states are located in different fixed points.

\vskip 0.5cm
\noindent
{\bf Target Space Picture}

We computed three and four-point functions 
at tree-level 
of closed $N=2$ strings on $\bC^2/\bZ_2$ and $T^4/\bZ_2$ orbifolds.
These amplitudes can be encoded
in a low-energy effective action. It is of the same form 
as that of the unorbifolded $N=2$ string \cite{Ooguri:1991fp}, with an additional
interaction term $ttu$.
We have not ruled out 
a possible four twisted states contact term $tttt$.
In addition to the usual on shell condition $p^2=0$ ($p_L^2=p_R^2=0$ for $T^4/\bZ_2$), one imposes the 
condition that every state is $\bZ_2$ invariant.  
Thus, for the $\bC^2/\bZ_2$ orbifold 
the on shell untwisted states are of the form ${1\over 2}(e^{ipX}+e^{-ipX})$ with $p^2=0$.

The low-energy effective action thus takes the form
\begin{equation}
S=\int\left(\partial \phi \bar{\partial}\phi
+{1\over 3}\, \phi \, \partial\bar{\partial} \phi \wedge
\partial\bar{\partial}\phi \, + tt\phi + {\cal O}(tttt)\right)
\end{equation}
($\phi$ representing an untwisted state and $t$ a twisted one)
where the condition that $\phi$ is $\bZ_2$ invariant is imposed on the external states.
Also note that the target-space is smaller now ($X^i$ identified with $-X^i$).

\subsection{Summary}

We have shown the existence of a single twisted state
in the $\bC^2/\bZ_2$ and $T^4/\bZ_2$ orbifolds, and computed 
the low-energy effective action. 
Unlike the $N=1$ superstring orbifold, 
in the $N=2$ superstring orbifold 
there is only a single twisted state, so there is only a single (real) modulus. 
It would be interesting to explore the behavior of this modulus 
by computing the form of the twist potential term $V(t)$ in the 
low-energy effective action.

\section{D-branes in The $N=2$ Orbifold}

The consistent D-branes of closed $N=2$ strings are of the types
$(0+0)$, $(2+0)$, $(0+2)$ and $(2+2)$ branes in $(2,2)$ signature
and  0-branes, 2-branes and 4-branes in $(4,0)$ signature
(see \cite{Gluck:2003pa} for details and \cite{JS:01} for an earlier work on D-branes in the unorbifolded $N=2$ string).
In this section we calculate 
scattering amplitudes of the orbifold theory in the presence of a D-brane  
located at a fixed point. We then consider the world volume effective theory
and compare it with the unorbifolded case.
Most of the calculations details which are relevant for this section can be found in appendices F and G.

\subsection{Two Twisted States Scattering Off D-branes}

We begin by considering twist fields in open strings.
Using this, the $XX$ two-point function in the presence of two twist fields on the disk is found
according to its behavior under the conformal symmetries.
Finally, the matter part of the two-point function is deduced in a similar method to that 
used in \cite{Dixon:1987} for the closed bosonic string.

In the closed string the OPEs of the bosonic field with the twist field take the form
(omitting index $i$)
\begin{eqnarray}
\del X(z)\sigma(w,\bar{w}) \sim (z-w)^{-1/2} \tau (w,\bar{w}) &,&
\bar{\del} X(\bar{z})\sigma(w,\bar{w}) \sim (\bar{z}-\bar{w})^{-1/2} \bar{\tau} (w,\bar{w}) \ ,
\label{clBOPE}
\end{eqnarray}
where $\tau$, $\bar{\tau}$ are excited twist fields.
Consider open strings on the upper half plane, ${\rm Im}z\ge 0$,
with the boundary conditions 
$\del X(z)=D \bar{\del}X(\bar{z})$ ($D=\pm1$) at $z=\bar{z}$. 
Close to the worldsheet boundary, (\ref{clBOPE}) 
must be corrected by a boundary term. 
After this correction the OPE takes the form
\begin{eqnarray}
\del X(z)\sigma(w,\bar{w}) &\sim& (z-w)^{-1/2}(z-\bar{w})^{-1/2}
(w-\bar{w})^{+1/2}\,\tau (w,\bar{w})  \ , \nn
\bar{\del} X(\bar{z})\sigma(w,\bar{w}) &\sim& D(\bar{z}-\bar{w})^{-1/2}
(\bar{z}-w)^{-1/2}(w-\bar{w})^{+1/2}\, \tau(w,\bar{w})  \ .
\end{eqnarray}

For every complex dimension we may thus write
\begin{eqnarray}
-{1\over 2}{\langle \partial_z X(z) \partial_w \bar{X}(w)\sigma(z_1,\bar{z_1})\sigma(z_2,\bar{z_2}) \rangle_{D_2} \over 
\langle \sigma(z_1,\bar{z_1})\sigma(z_2,\bar{z_2}) \rangle_{D_2}} &=& \nn
(z-z_1)^{-1/2} (z-z_2)^{-1/2} (z-\bar{z}_1)^{-1/2} (z-\bar{z}_2)^{-1/2} &\times& \nn
(w-z_1)^{-1/2} (w-z_2)^{-1/2} (w-\bar{z}_1)^{-1/2} (w-\bar{z}_2)^{-1/2} 
&\times& g(z,w,z_i,\bar{z}_i) \ , 
\label{XXTT}
\end{eqnarray}
where the function $g(z,w,z_i,\bar{z}_i)$ is single-valued and is holomorphic in $z$ and $w$.
Its form can be deduced from the behavior of 
(\ref{XXTT}) as $z\rightarrow w$ and under the conformal symmetries of the disk.

By subtracting $1/(z-w)^2$ from (\ref{XXTT}) and taking the limit $z\rightarrow w$ we get the 
normalized energy-momentum tensor expectation value
$\langle T(z)\sigma(z_1,\bar{z_1})\sigma(z_2,\bar{z_2}) \rangle_{D_2} /
\langle \sigma(z_1,\bar{z_1})\sigma(z_2,\bar{z_2}) \rangle_{D_2}$.
In the limit $z\rightarrow z_1$, the coefficient of its $(z-z_1)^{-1}$ term is 
$\partial_{z_1} \ln \langle \sigma(z_1,\bar{z_1})\sigma(z_2,\bar{z_2})\rangle_{D_2}$.
Repeating this procedure for the antiholomorphic part of the energy-momentum tensor and integrating, gives
the twist fields two-point function on the disk.
we get
\begin{equation}
\langle \sigma(z_1,\bar{z_1})\sigma(z_2,\bar{z_2})\rangle_{D_2} = 
(z_1-\bar{z_1})^{A+B-{1\over 12}}(z_1-z_2)^{-B-{1\over 12}}
(z_1-\bar{z_2})^{-A-{1\over 12}}\, \times\, ({\mbox c.c.}) \ .
\label{2sigD}
\end{equation}
Where $A$, $B$ are yet unknown constants.

Looking at the four twisted states correlation function on $S^2$ (\ref{Poisson}), we 
see that the OPE of two twist fields is of the form (ignoring possible boundary terms):
\begin{equation}
\sigma(z_1,\bar{z_1})\sigma(z_2,\bar{z_2})\sim (z_1-z_2)^{-1/4}(\bar{z}_1-\bar{z}_2)^{-1/4} ...
\end{equation}
Therefor in (\ref{2sigD}) one must have $B={1\over 6}$.

The two twisted states correlation function in the presence of a D-brane is
\begin{eqnarray}
\cA_{ttD}&=&\int dz_1 \, \langle \sigma^1\sigma^2(z_1,\bar{z_1})\sigma^1\sigma^2(z_2,\bar{z}_2)\rangle_{D_2}
\cdot\langle e^{-\phi^+/2}(z_1)e^{-\phi^+/2}(z_2)e^{-\bar{\phi}^+/2}(\bar{z}_1)e^{-\bar{\phi}^+/2}(\bar{z}_2)\rangle_{D_2}  \nn
&\cdot&\langle e^{-\phi^-/2}(z_1)e^{-\phi^-/2}(z_2)e^{-\bar{\phi}^-/2}(\bar{z}_1)e^{-\bar{\phi}^-/2}(\bar{z}_2)\rangle_{D_2} 
\cdot\langle c(z_2)\bar{c}(\bar{z}_1)\bar{c}(\bar{z}_2)\rangle_{D_2} \nn
&=& \int  dz_1 \, (z_1-\bar{z}_1)^{4A-{1\over 6}}(z_1-z_2)^{-1} (z_1-\bar{z}_2)^{-2A-{2\over 3}}
(\bar{z}_1-z_2)^{-2A+{1\over 3}}(z_2-\bar{z}_2)^{1\over 2} \ .
\end{eqnarray}
This should
 be invariant under the coordinate transformation $z\rightarrow -1/z$, which gives $A=-{1\over 12}$.
In fact, this is also necessary in order to ensure that (\ref{2sigD}) is invariant under the exchange of the twist fields
$(z_1,\bar{z}_1)\leftrightarrow (z_2,\bar{z}_2)$.
By substituting $A=-{1\over 12}$ into (\ref{2sigD}) we see that the open string OPE of two twist fields near the worldsheet boundary includes no boundary terms.
On the disk we may fix $z_2=i$, $z_1=iy$ where $y$ is a real number between $0$ and $1$. Thus,
\begin{eqnarray}
\cA_{ttD}&=&\int_0^1  dy \, y^{-1/2} (1-y)^{-1} \ .
\end{eqnarray}

The integral diverges. The divergence comes from the area of integration $y\rightarrow 1$, where the two 
twist fields are close to each other.
Note that the leading term of the integrand at this regime is $(1-y)^{-1}$. 
This should be expected, because the OPE of the two twisted states located
at $iy$ and $i$ should be of the form $(1-y)^{-1+p^2}$ where $p$ is the
momentum of the intermediate state; $p=0$ contributes the leading term.

We propose the following low-energy effective action picture:
The diagram contributing to $\cA_{ttD}$ is a combination of a $ttu$ diagram,
and a coupling term of $u$ to the D-brane, 
with a propagator $uu$ connecting them. 
The coupling of a single $u$ to a D-brane is computed in
\cite{Gluck:2003pa} and found to be given by a constant.
Recall also that the off-shell $ttu$ vertex assumed to be constant as well.
Hence the diagram is proportional to
$\int d^{4-p}k_{\perp}\,1/k_{\perp}^2$, where $k_{\perp}$
is the momentum of the intermediate untwisted state
normal to the D$p$-brane world volume.
This is because both the $ttu$ vertex and the D-brane break
translation invariance along the normal directions of the
D-brane.
We now interpret that this divergent integral is responsible for
the divergent result of $\cA_{ttD}$.

\vskip 0.3cm
\noindent
{\bf Other Amplitudes}

Scattering amplitudes of untwisted states 
in the presence of D-branes are identical to those found in 
\cite{Gluck:2003pa} \footnote{Some of these amplitudes were previously computed in \cite{JS:01}.}, with 
the appropriate $\bZ_2$ invariance taken into account. 
Thus, instead of each string of momentum $p$, one sums up 
the amplitudes for a string of momentum $p$ and the amplitudes 
for a string of momentum $-p$.
We find that all the amplitudes are
unchanged.
The only exception is  the three open string untwisted states amplitude, which is 
zero in the orbifold theory.

\subsection{Effective World Volume Theory}

\vskip 0.3cm
\noindent
{\bf The $(2+2)$ Brane}

In the unorbifolded theory, the effective theory in the case of the $(2+2)$ (space filling) brane 
is self-dual Yang-Mills coupled to gravity \cite{mar;92}. The self-dual YM 
equation reads 
\begin{equation}
F_{IJ}={1\over 2}\epsilon_{IJKL}F^{KL} \ ,
\label{SDYMeom}
\end{equation}
where
\begin{equation}
F_{IJ}=-[D_I,D_J],~~D_I=\partial_I-A_I \ .
\end{equation}
and (in complex coordinates)
\begin{eqnarray}
A_i \equiv e^{-i\varphi/2}\del_i e^{i\varphi/2} &,& 
\bar{A}_{\bar{i}} \equiv e^{i\varphi/2}\bar{\del}_{\bar{i}} e^{-i\varphi/2}
\label{SDYMA}
\end{eqnarray}
with $\varphi$ being the open string state.

In the orbifold theory, the only difference is that now the closed string state is coupled to the twisted state. Thus, 
the effective action should be the same as in \cite{mar;92} but with the additional $\phi tt$ term (and possible terms 
of order $t^4$). It reads
\begin{equation}
S=\int\left(\partial \phi \bar{\partial}\phi
+{1\over 3}\, \phi \, \partial\bar{\partial} \phi \wedge
\partial\bar{\partial}\phi \, + tt\phi + {\cal O}(t^4)
-Tr(\partial \varphi \bar{\partial}\varphi) + \partial \bar{\partial}\phi \wedge Tr(\varphi\partial\bar{\partial}\varphi)
+ {\cal O}(\varphi^3)
\right)
\end{equation}
where 
$\bZ_2$ acts on both the target space indices and the Chan-Paton
indices, yielding the quiver-type gauge theories \cite{Douglas:1996sw}.

\vskip 0.3cm
\noindent
{\bf Lower Dimensional Branes}

In lower dimensional branes the unorbifolded world volume theory is a dimensional 
reduction of self-dual Yang-Mills \cite{Gluck:2003pa}. However,
 due to (\ref{SDYMA}) $A_i$ vanish
in all directions perpendicular to the brane, so that there are no
worldvolume fields that correspond 
to these directions.
For the $(2+0)$-brane the equations of motion now read \cite{mar;92}
\begin{eqnarray}
\bar{\del} \left(e^{-i\varphi/2}\del e^{i\varphi/2}\right) = 0 &,& 
\del\left(e^{i\varphi/2}\bar{\del} e^{-i\varphi/2}\right) = 0  \ ,
\end{eqnarray}
giving a WZW theory.
Closed - open strings scatterring amplitudes vanish, 
while closed string one-point function on $D_2$ is non-vanishing \cite{Gluck:2003pa}. 
Thus, for the $(2+0)$-brane the effective world-volume theory is a 
two-dimensional WZW theory with only one 
coupling to the bulk
given by the one-point function.

In the orbifolded theory, we have seen no coupling of open states to twisted states up to the level 
considered (namely, no twisted-twisted-open pole appeared in $\cA_{ttD}$).
Therefor the effective world-volume theory is again a two-dimensional WZW theory, the only difference being the 
target-space identification under $\bZ_2$ (which changes the world-volume topology as well).

In the $(0+0)$-brane there are no degrees of freedom as $A_i$ vanishes
in all directions, and the world-volume theory is trivial.

\subsection{Summary}

We have computed twist
 correlators in the presence of
D-branes and discussed the world volume effective action.
The twisted states exhibit 
infinite scattering amplitudes off D-branes. We leave the further investigation of this to future work.

\section*{Acknowledgments} 

We would like to thank O. Aharony and D. Kutasov for discussions.
T.S. would like to thank Tel Aviv University, where most parts of this
work have been done.

\newpage

\appendix

\section{Free Fields Representation of the $N=2$ Superconformal Algebra}

The $N=2$ superconformal algebra is
\begin{eqnarray}
\left[L_m , L_n\right] &=& (m-n)L_{m+n} + {c\over 12} m(m^2-1) \delta_{m+n} \nn
\left[J_m , J_n\right] &=& {c\over 3} m \delta_{m+n} \nn
\left[L_n,J_m\right] &=& -mJ_{m+n} \nn
\left[L_n,G^{\pm}_r\right] &=& \left({n\over 2}-r)\right)G^{\pm}_{n+r} \nn
\left[J_n,G^{\pm}_r\right] &=& \pm G^{\pm}_{n+r} \nn
\{G^{\pm}_r,G^{\pm}_s\} &=& 0 \nn
\{G^+_r,G^-_s\} &=& 2L_{r+s} + (r-s)J_{r+s} + {c\over 3}\left(r^2-{1\over 4}\right)\delta_{r+s}
\end{eqnarray}

The free fields representation of this algebra reads
\begin{eqnarray}
L^M_n &=& {1\over 2} \alpha^i_k\bar{\alpha}_{i~n-k} + {1\over 4}(2t-n)\bar{\psi}^i_{n-t}\psi_{i~t} + a^M\delta_n \, ,\nn
L^g_n &=& (2n-m)b_m c_{n-m} + (n-m)\hat{b}_m\hat{c}_{n-m}
+ ({3\over 2}n-s)\beta^+_s\gamma^-_{n-s} + ({3\over 2}n-s)\beta^-_s\gamma^+_{n-s} + a^g\delta_n \, ,\nn
G^{M+}_r &=& {1\over \sqrt{2}}\bar{\alpha}^i_k\psi_{i~r-k} \, \, , \, \,
G^{M-}_r = {1\over \sqrt{2}}\alpha^i_k\bar{\psi}_{i~r-k} \nn
G^{g\pm}_r &=& -{1\over 2}(2r+n)\beta^\pm_{r-n}c_n \pm \hat{c}_n\beta^\pm_{r-n}
-2b_n\gamma^{\pm}_{r-n} \mp (2r-n)\hat{b}_n\gamma^\pm_{r-n} \, ,\nn
J^M_n &=& {1\over 2} \psi^i_t\bar{\psi}_{i~n-t} + a_J^M\delta_n \, , \nn
J^g_n &=& \gamma^-_r\beta^+_{n-r} - \gamma^+_r\beta^-_{n-r} - nc_m\hat{b}_{n-m} + a_J^g\delta_n \, .
\end{eqnarray}
where $n,m\in\bZ$, $k\in\bZ+\phi_0$, $t\in\bZ+{1\over2}+\phi+\phi_0$, $r,s\in\bZ+{1\over2}\pm\phi$.

And the BRST charge is
\begin{eqnarray}
Q_{BRST} &=& :(L^M_n + {1\over 2} L^g_n + {a^g\over 2})c_{-n}: + :(J^M_n + {1\over 2} J^g_n+ {a_J^g\over 2})\hat{c}_{-n}: \nn
&+& :(G^{M+}_r + {1\over 2} G^{g+}_r)\gamma^-_{-r}: + :(G^{M-}_r + {1\over 2} G^{g-}_r)\gamma^+_{-r}:
\end{eqnarray}

So that 
\begin{eqnarray}
L_n &=& \{Q_{BRST},b_n\}\nn
J_n &=& \{Q_{BRST},\hat{b}_n\}\nn
G^{\pm}_r &=& [Q_{BRST},\beta^{\pm}]
\end{eqnarray}

\section{Partition function}

\subsection{Moduli Integration Measure}

\subsubsection{The $U(1)_V\times U(1)_A$ Gauge Transformations}

A Polyakov-type worldsheet  description of the $N=2$ string is given by 
the action of two-dimensional $(2,2)$ supergravity.  
The fields are a graviton, two gravitini and a vector field,
interacting with complex scalars  $X^i$, and Dirac spinor fermions  
$\Psi^i=(\psi^i_L,\psi^i_R)$, with $i=1,2$.
The gravitini can be  set to zero 
by the supersymmetry transformations, and  in the following we 
will consider vector field configurations
 where the  worldsheet instanton number is zero
(see e.g. \cite{allloop} for instanton numbers different than zero).
We will be working with  Euclidean worldsheets.
 The two-dimensional supergravity action reads then

\begin{equation}
S = {1\over 2\pi} \int d^2\sigma g^{1/2} \left({1\over 2\alpha^\prime} \partial_a X^i \partial^a \bar{X_i} + 
{i\over 2} {\Psi^i}^\dagger\Gamma^a\partial_a\Psi_i + A_a{\Psi^i}^\dagger\Gamma^a\Psi_i \right)
\  , 
\label{action}
\end{equation}
where  $a=1,2$, $A_a$ is a real vector field, $\Gamma^a = \sigma^a$ and
\begin{eqnarray}
{\psi^i_L}^\dagger = {\bar \psi}^i_R &,& {\psi^i_R}^\dagger = {\bar \psi}^i_L
\label{ferhc} \ .
\end{eqnarray}
The index $i$ in $X^i$, $\psi_{L,R}^i$ ($\bar{X}^i$, $\bar{\psi}_{L,R}^i$) is understood 
as a holomorphic (anti-holomorphic) index of the target space.

It was pointed out in \cite{FT:1981} (in the Lorentzian case) that the two-dimensional supergravity action has  
both axial and vector gauge symmetries, and that they are both required in the gauge
fixing procedure.
Consider the vector and axial gauge symmetries, whose transformations are parameterized by the real parameters $\alpha_V$ 
and $\alpha_A$,
\begin{eqnarray}
A_a \rightarrow A_a+{1\over 2}(\partial_a\alpha_V + \epsilon_{ab}\partial^b\alpha_A) &,& 
\Psi^i \rightarrow e^{i\alpha_V + \sigma^3\alpha_A} \Psi^i \  ,
\end{eqnarray}
 $\epsilon_{12}=1$.
Defining  $w=\sigma_1+i\sigma_2$,  
\begin{eqnarray}
A \equiv A^w \rightarrow A^w+\bar{\partial}\alpha &,&
\bar{A} \equiv A^{\bar{w}} \rightarrow A^{\bar{w}}+\partial\bar{\alpha} \nn
\psi^i_L \rightarrow e^{i\alpha} \psi^i_L &,& \psi^i_R \rightarrow e^{i\bar{\alpha}} \psi^i_R \nn
{\bar \psi}^i_L \rightarrow e^{-i\alpha} {\bar \psi}^i_L &,&
{\bar \psi}^i_R \rightarrow e^{-i\bar{\alpha}} {\bar \psi}^i_R \ ,
\label{gaugetrans}
\end{eqnarray}
where $\alpha \equiv \alpha_V - i\alpha_A$.
Note that $\bar{A} = A^\dagger$.

The generalization of  (\ref{gaugetrans}) 
for a field 
${\cal O}$  with $U(1)$ charges 
$J_0$, $\bar{J_0}$ with  
respect to the gauge fields $A$, $\bar{A}$ reads

\begin{equation}
{\cal O} \rightarrow e^{i(\alpha J_0 + \bar{\alpha}\bar{J_0})} {\cal
  O} \ .
\end{equation}

Here $\psi^i_{L,R}$ may be periodic up to the phases $2\pi\phi_{L,R}$, which is more general 
than the allowed periodicity of the $N=1$ superstring.
In this case ${\bar \psi}^i_{L,R}$ are periodic up to the phases $-2\pi\phi_{L,R}$. 
However, (\ref{ferhc}) implies  $\phi^\prime \equiv \phi_L=\phi_R$.
This is consistent with the $N=2$ algebra only if the supercurrents, the superghosts and 
anti-superghosts have the same periodicity as the fermions of the same $U(1)$ charges.
Thus,  
every field ${\cal O}$ satisfies
\begin{equation}
{\cal O} (w+2\pi, \bar{w}+2\pi) = e^{2\pi i(J_0 + \bar{J_0})\phi^\prime} {\cal O} (w, \bar{w}) \ ,
\label{phiprime}
\end{equation}
with the same $\phi^\prime$ for all fields.


\subsubsection{Moduli on $T^2$ and Their Measure}

The vector field 
$A$ is a two-component field, which may be set to zero  by the two gauge transformations.
However, we must consider possible Killing vectors, i.e. gauge transformations which do 
not change $A$, and moduli 
 which are variations of $A$ orthogonal to the gauge transformations.

The Killing vectors are found by solving the equation
\begin{equation}
\partial_a\alpha_V + \epsilon_{ab}\partial^b\alpha_A = 0 \ ,
\end{equation}
or equivalently
\begin{equation}
\partial\alpha = 0 \ .
\end{equation}
On a compact Riemann surface, the only solution for this 
equation is $\alpha={\rm const.}$, and  for all closed topologies, 
there is one complex (or two real) Killing vectors.
Therefor, there will be one insertion of $\hat{c}$ (the $U(1)$ ghost), 
and one insertion of $\bar{\hat{c}}$ in every amplitude.
Since the Killing vectors are 
constant, these insertions do not affect the calculations. 

The moduli are found by solving:
\beq
\int \, d^2\sigma \, g^{1/2} \delta A^a (\partial_a\alpha_V + 
\epsilon_{ab}\partial^b\alpha_A) = 0 \ , 
\eeq
or similarly
\beq
\int \, d^2\sigma \, g^{1/2} (\nabla_a (\delta A^a) \alpha_V + 
\epsilon_{ab}\nabla^b (\delta A^a)\alpha_A) 
= 0 \ .
\eeq
Thus, we have
$\nabla_a (\delta A^a) = 0$, $\epsilon_{ab}\nabla^b (\delta A^a) = 0$  or equivalently
\begin{eqnarray}
\partial(\delta A) = 0 &,& \bar{\partial}(\delta \bar{A}) = 0 \ .
\label{moduli}
\end{eqnarray}
These equations have no
solution on $S^2$,  and therefor $A,\bar{A}$ can be set to zero on the 2-sphere.

On $T^2$, $A$ can be gauge transformed to a complex constant.
This  can be seen 
directly by expanding $A$, which is 
a doubly periodic function, in a sine and cosine series and constructing explicitly the
gauge transformation function 
\footnote{$A$ will be expanded in $\cos(a_{m,n}w+\bar{a}_{m,n}\bar{w})$ and 
$\sin(a_{m,n}w+\bar{a}_{m,n}\bar{w})$ where $a_{m,n} \equiv {i\over 2\tau_2}(m\bar{\tau}+n)$.}.
Alternatively,
a solution to (\ref{moduli}) is given by a holomorphic one
form of $T^2$ is thus a (complex) constant. 
We will therefor have one complex modulus. 
The path integral on the torus worldsheet includes a sum over 
all configurations of the gauge fields,
which are the constants after gauge fixing.

The fermions obey the equations of motion:
\begin{eqnarray}
\bar{\partial}\psi_L = iA\psi_L &,& \partial\psi_R = i\bar{A}\psi_R \ ,
\end{eqnarray}
from which it follows that
\begin{eqnarray}
\psi_L(w,\bar{w}) = e^{i (A\bar{w}+\bar{A}w)}\psi^\prime_L(w) &,& 
\psi_R(w,\bar{w}) = e^{i (A\bar{w}+\bar{A}w)}\psi^\prime_R(\bar{w}) \ ,
\end{eqnarray}
where $\psi^\prime_L(w)$ is holomorphic and $\psi^\prime_R(\bar{w})$ is antiholomorphic. 
Since $\psi_{L,R}(w,\bar{w})$ are truly doubly periodic,
the fields $\psi^\prime_{L,R}$ are doubly periodic only up to 
the phase shifts
\begin{eqnarray}
\psi^\prime_L(w+2\pi\tau) = e^{- 2i\pi (A\bar{\tau}+\bar{A}\tau)} \psi^\prime_L(w) &,&
\psi^\prime_L(w+2\pi) = e^{- 2i\pi (A+\bar{A})} \psi^\prime_L(w) \ ,
\end{eqnarray}
with the same phase shifts for $\psi^\prime_R$ as well.
We denote the phase shifts by
\begin{eqnarray}
\theta = -(\bar{\tau}A+\tau\bar{A}) &,& \phi = -(A+\bar{A}) \ .
\end{eqnarray}
It will later be useful to define 
$u\equiv -\theta+\phi\tau$, so that $A = i(\phi\tau-\theta)/2\tau_2 = iu/2\tau_2$.

The same holds for any field with a non-zero $U(1)$ charge
(including the superghosts).  
They can be written as a holomorphic (antiholomorphic) field multiplied by a Wilson line, with the 
holomorphic (antiholomorphic) field being doubly periodic only up to phase shifts
\begin{eqnarray}
{\cal O}(w+2\pi\tau,\bar{w}+2\pi\bar{\tau}) = e^{2\pi i(J_0+\bar{J}_0)\theta}{\cal O}(w,\bar{w}) &,& 
{\cal O}(w+2\pi,\bar{w}+2\pi) = e^{2\pi i(J_0+\bar{J}_0)\phi}{\cal O}(w,\bar{w}) \ . \nn
\label{phase}
\end{eqnarray}
Note that any phase shift $\phi^\prime$ as in (\ref{phiprime}) can be 
absorbed in $\phi$ by a redefinition 
of the gauge field.

The partition function on the torus is defined by tracing over all 
oscillators on the unit circle (the $\alpha$-cycle) shifted by
$2\pi\tau$,
with the non-trivial periodicities (\ref{phase}).
That along the $\alpha$-cycle 
(parametrized by the phase shift $2\pi\phi$) affects the conformal 
weight of these operators.
One has to insert the operator $e^{-2 \pi i \theta (J_0+\bar{J_0})}$
into the trace in order to impose the boundary condition in
(\ref{phase}) along the $\beta$-cycle.

Another approach is to 
add ${1\over 2\pi} \int d^2\sigma g^{1/2} [-2(AJ+\bar{A}\bar{J})] = 2\pi (uJ_0+\bar{u}\bar{J}_0)$ to the action. These
two approaches are equivalent, as is implied by the equality (to be confirmed later)
\begin{equation}
d^2\tau d\theta\, \trphi \left(q^{L_0}\bar{q}^{\bar{L}_0} e^{-2\pi i\theta (J_0+\bar{J}_0)}\right) = 
d^2\tau d\theta d\phi\, \tr \left(q^{L_0}\bar{q}^{\bar{L}_0} 
e^{2\pi i (u J_0 + \bar{u}\bar{J}_0)}\right) \ .
\label{u1dep}
\end{equation}

Finally we should compute the integration measure. Tracing over $\phi$ and $\theta$ replaces 
the 
tracing over the gauge field. 
There is one complex modulus $u$ for the gauge field $A$. We get
\begin{eqnarray}
\partial_u A = {i\over 2\tau_2}  &,& \partial_{\bar{u}}\bar{A} = -{i\over 2\tau_2} \ .
\end{eqnarray}
The integration measures can now be calculated from the ghost and antighost insertions
\begin{eqnarray}
&&{1\over 4\pi}\left(\hat{b},\partial_u A\right) \hat{c} = 
{i \over 8\pi\tau_2} \int d^2 \sigma g^{1/2} \hat{b} \hat{c}
 = {\pi\over 2} \hat{b}_0 \hat{c}_0  \ , \nn 
&&{1\over 4\pi}\left(\bar{\hat{b}},\partial_{\bar{u}} \bar{A}\right) \bar{\hat{c}} = 
{-i \over 8\pi\tau_2} \int d^2 \sigma g^{1/2} \bar{\hat{b}}
 \bar{\hat{c}} = {\pi\over 2} \bar{\hat{b}}_0 \bar{\hat{c}}_0  \ .  
\end{eqnarray}

Note that the torus amplitude contains two $U(1)$ anti-ghost insertions 
for the one complex modulus.
The total integration measure for all moduli reads
\begin{equation}
{d^2\tau\over 4\tau_2} du d\bar{u}{1\over 4\pi}\left(\hat{b},\partial_u A\right) \hat{c} 
{1\over 4\pi}\left(\bar{\hat{b}},\partial_{\bar{u}} \bar{A}\right) \bar{\hat{c}} =  
{\pi^2\over 8} d^2\tau d\theta d\phi \hat{b}_0 \hat{c}_0 \bar{\hat{b}}_0 \bar{\hat{c}}_0 \ .
\label{measure}
\end{equation}

\subsection{Oscillators}

\subsubsection{Bosons}

Working with the complex coordinates $z$, $\bar{z}$ with $z=e^{-iw}$, we may Laurent expand 
the complex scalars:
\begin{equation}
\del X^i(z)=-i \sqrt{\alpr\over 2}\,\sum_r {\alpha^i_r\over z^{r+1}} \ ,
~~~
\del \bar{X^i}(z)=-i \sqrt{\alpr\over 2}\,\sum_s {\bar{\alpha}^i_s\over z^{s+1}} \ .
\end{equation}
Here
\begin{equation}
r\in \bZ+\phi_0 \ ,~~~s\in \bZ-\phi_0 \ ,
\end{equation}
where $0\le\phi_0<1$ gives a twist. For a $\bZ_N$ twist, 
$\phi_0=k/N,~k=0,1,\cdots, N-1.$
Similar expansions in $\bar{z}$ hold for $\bar{\del}X^i$ and $\bar{\del}\bar{X}^i$, but with 
$\phi_0\rightarrow -\phi_0$ (this is due to the fact that as $w\rightarrow w+2\pi$, $z\rightarrow e^{-2\pi i}z$ while 
$\bar{z}\rightarrow e^{2\pi i}\bar{z}$).

The commutation relations read
\begin{equation}
[\alpha^i_r,\bar{\alpha}^j_s]=2r\,\delta_{r+s}\, \eta^{ij} \ ,
\end{equation}
and the Virasoro generators take the form
\begin{equation}
L_n={1\over 2}\sum_{r\in\bZ+\phi_0} : 
\alpha^i_r\bar{\alpha}_{i\,n-r}:
\,+\,a\,\delta_n \ .
\end{equation}
The ground state $\gs_{X}$ is defined by
\begin{equation}
\alpha^i_r \gs_{X}=\bar{\alpha}^i_r \gs_{X}=0 \ , ~{\rm for}~r> 0 \ .
\end{equation}
Note that for $\phi_0=0$, there is a set of states $|0;p\rangle_X$ defined by their (complex) eigenvalues under 
$\alpha^i_0$ (which are the complex conjugates of the eigenvalues under 
$\bar{\alpha}^i_0$). We denote these eigenvalues by  
$\sqrt{\alpr\over 2}p^i_L$ and $\sqrt{\alpr\over 2}p^i_R$ for 
the holomorphic and anti-holomorphic oscillators, 
respectively.

For every complex scalar, $a$ can be evaluated in the $w$ coordinates
 using a zeta function regularization:
\begin{equation}
a\!=\! {1\over 2}\sum_{n=0,1,\cdots}\!(n+\phi_0) +{1\over 2}\sum_{n=1,2,\cdots}\!(n-\phi_0)\!=\! 
-{1\over 12}+ {1\over 2}\phi_0(1-\phi_0) \ .
\label{bosonzpe}
\end{equation}

For the computation of the $N=2$ string partition function we 
consider two complex scalars, and each is expanded in 
both holomorphic modes and anti-holomorphic modes.
For the $\bZ_2$ orbifold, the projection operator is given by
$\cP=(1+{\bf r})/2$ where ${\bf r}$ is the $\bZ_2$ reflection.
In the untwisted sector $\phi_0=0$, and one finds that in the compact case $T^4/\bZ_2$,
\begin{eqnarray}
\trphi \left( q^{L_0}\bar{q}^{\bar{L}_0} e^{-2\pi i\theta (J_0+\bar{J}_0)}\right) 
&\!=\!&
\sum_{p\in\Gamma} \,\langle p|p \rangle 
\,q^{{\alpha^{\prime}\over 4}p_L^2-{1\over 6}}
\bar{q}^{{\alpha^{\prime}\over 4}p_R^2-{1\over 6}}
\prod_{n=1}^{\infty}\left|1-q^n\right|^{-8} \nn
&\!=\!&
\sum_{p\in\Gamma}\, q^{{\alpha^{\prime}\over 4}p_L^2} \bar{q}^{{\alpha^{\prime}\over 4}p_R^2}
\left|\eta(\tau)\right|^{-8} \ ,\nn
\trphi \left( {\bf r}\cdot q^{L_0}\bar{q}^{\bar{L}_0} e^{-2\pi i\theta (J_0+\bar{J}_0)}\right) 
&\!=\!&
\sum_{p\in\Gamma} \,\langle -p|p \rangle 
\,q^{{\alpha^{\prime}\over 4}p_L^2-{1\over 6}}
\bar{q}^{{\alpha^{\prime}\over 4}p_R^2-{1\over 6}}
\prod_{n=1}^{\infty}\left|1+q^n\right|^{-8} \nn
&\!=\!& 2^4\cdot {\left| \eta(\tau)\right|^4\over\left| \vthe_{10}(0,\tau)\right|^4} \ .
\end{eqnarray}

Where $p \equiv (p_L,p_R)$ and $\Gamma$ is an 8-dimensional lattice which is self-dual and even under the product 
$p\cdot p^\prime \equiv p_L\cdot p^\prime_L - p_R\cdot p^\prime_R$ 
(this is needed for modular invariance).

In the non-compact case $\bC^2/\bZ_2$, $p \equiv p_L=p_R$ and the sum 
is replaced by an integral over $p$.
$\langle -p|p \rangle = \delta^4 (p+p) = 2^{-4} \delta (p)$,
 $\langle p|p \rangle = \delta^4 (0) =(2\pi)^{-4}V$, and we get
\begin{eqnarray}
\trphi \left( q^{L_0}\bar{q}^{\bar{L}_0} e^{-2\pi i\theta (J_0+\bar{J}_0)}\right) 
&\!=\!&
\int d^4p \,\langle p|p \rangle 
\,(q\bar{q})^{{\alpha^{\prime}\over 4}p^2-{1\over 6}}
\prod_{n=1}^{\infty}\left|1-q^n\right|^{-8}
\!=\!
{V\over (4\pi^2\alpha^{\prime}\tau_2)^2}\left|\eta(\tau)\right|^{-8} \ ,\nn
\trphi \left( {\bf r}\cdot q^{L_0}\bar{q}^{\bar{L}_0} e^{-2\pi i\theta (J_0+\bar{J}_0)}\right) 
&\!=\!&
\int d^4p \,\langle -p|p \rangle 
\,(q\bar{q})^{{\alpha^{\prime}\over 4}p^2-{1\over 6}}
\prod_{n=1}^{\infty}\left|1+q^n\right|^{-8} 
\!=\! {\left| \eta(\tau)\right|^4\over\left| \vthe_{10}(0,\tau)\right|^4} \ .
\end{eqnarray}

For each twisted sector ($\phi_0=1/2$), one gets
\begin{eqnarray}
&& \trphi \left(q^{L_0}\bar{q}^{\bar{L}_0} e^{-2\pi i\theta (J_0+\bar{J}_0)}\right)
=(q\bar{q})^{1\over 12}\prod_{n=1}^{\infty}\left|1-q^{n-{1\over 2}}\right|^{-8} 
\!=\!
{\left| \eta(\tau)\right|^4\over\left| \vthe_{01}(0,\tau)\right|^4} \ , \nn
&& \trphi \left({\bf r}\cdot q^{L_0}\bar{q}^{\bar{L}_0} e^{-2\pi i\theta (J_0+\bar{J}_0)}\right)
=(q\bar{q})^{1\over 12}\prod_{n=1}^{\infty}\left|1+q^{n-{1\over 2}}\right|^{-8} 
\!=\!
{\left| \eta(\tau)\right|^4\over\left| \vthe_{00}(0,\tau)\right|^4} \ .
\end{eqnarray}
In a compact case there are $2^4$ fixed points, and therefor $2^4$ twisted sectors.
In the non-compact case there is only one twisted sector.

\subsubsection{Fermions}
Each holomorphic complex fermion can be Laurent expanded as follows
\begin{equation}
\psi^i_L(z)=\sum_r {\psi^i_r\over z^{r+1/2}} \ ,
~~~
\bar{\psi}^i_L(z)=\sum_s {\bar{\psi}^i_s\over z^{s+1/2}} \ .
\end{equation}
Here
\begin{equation}
r\in \bZ+{1\over2}+\tilde{\phi}\ ,~~~s\in \bZ+{1\over2}-\tilde{\phi}\ ,
\end{equation}
where $\tilde{\phi} \equiv \phi+\phi_0$.
$\phi$ is equivalent to $\phi+1$ and we may therefor assume $-1/2\le\phi <1/2$.
A similar expansion in $\bar{z}$ holds for $\psi^i_R$ and $\bar{\psi}^i_R$, but with 
$\tilde{\phi} \rightarrow -\tilde{\phi}$ (due to the same reason as in the bosonic fields).

The anti-commutation relations are
\begin{equation}
\{ \psi^i_r,\bar{\psi}^j_s\}=2\delta_{r+s}\,\eta^{ij} \ ,
\end{equation}
and the Virasoro generators take the form
\begin{equation}
L_n={1\over 4}\sum_r(2r-n):
\bar{\psi}^i_{n-r}\psi_{i\,r}:
+a\,\delta_n \ .
\end{equation}
The ground state $\gs_{\psi}$ is defined by
\begin{equation}
\psi^i_r \gs_{\psi}= \bar{\psi}^i_r \gs_{\psi}=0 \ , ~{\rm for}~r> 0 \ .
\end{equation}

Note that for a half-integer $\tilde{\phi}$ there will be $2$ different ground states for every complex fermion $\psi^i$, 
one annihilated by $\psi^i_0$ and the other by $\bar{\psi}^i_0$. 
These operators may also be used to move from one ground state to the other.
Totally there are $2^D\cdot 2^D$ such states, for $D$ holomorphic and $D$ antiholomorphic complex fermions. 
All of them appear in the partition function.

We will  calculate the value of $a$ in the $w$ coordinate separately for two cases:

case (i): $0\le \tilde{\phi}+{1\over 2}<1$. Then for every holomorphic complex fermion:
\begin{eqnarray}
a&\!=\!& -{1\over 2}\sum_{n=0,1,\cdots}\! \left(n+{1\over 2}+\tilde{\phi}\right)
- {1\over 2}\sum_{n=1,2,\cdots}\! \left(n-{1\over 2}-\tilde{\phi}\right) 
\!=\!  -{1\over 24}+ {1\over 2}\,\tilde{\phi}^2 \ .
\label{fermionzpe1}
\end{eqnarray}

case (ii): $1\le \tilde{\phi}+{1\over 2}<2$. Then for every holomorphic complex fermion:
\begin{eqnarray}
a&\!=\!&  -{1\over 2}\sum_{n=-1,0,\cdots}\! \left(n+{1\over 2}+\tilde{\phi}\right)
- {1\over 2}\sum_{n=2,3,\cdots}\! \left(n-{1\over 2}-\tilde{\phi}\right) 
\!=\!  -{1\over 24}+ {1\over 2}\,(\tilde{\phi}-1)^2 \ .
\label{fermionzpe2}
\end{eqnarray}

For the computation of the $N=2$ string partition function, we take
two holomorphic complex fermions and two antiholomorphic 
complex fermions. 
In  the untwisted sector ($\phi_0=0$), one may
use the case (i), which is consistent with $-1/2\le\phi< 1/2$.
One gets 
\begin{eqnarray}
&&\trphi \left(q^{L_0}\bar{q}^{\bar{L}_0} e^{-2\pi i\theta (J_0+\bar{J}_0)}\right) 
= \nn
&&=(q\bar{q})^{-{1\over 12}+\phi^2} 
 \cdot  \prod_{n=1}^{\infty}
(1+q^{n-{1\over 2}+\phi}e^{-2\pi i\theta})^2
(1+q^{n-{1\over 2}-\phi}e^{2\pi i\theta})^2
(1+\bar{q}^{n-{1\over 2}-\phi}e^{-2\pi i\theta})^2
(1+\bar{q}^{n-{1\over 2}+\phi}e^{2\pi i\theta})^2 
 \nn
&&= {1\over\left|\eta(\tau)\right|^4} \left|\cthe{\phi}{-\theta}(0,\tau)\right|^4 \ , \nn
\nn
&&\trphi \left({\bf r}\cdot q^{L_0}\bar{q}^{\bar{L}_0} e^{-2\pi i\theta (J_0+\bar{J}_0)}\right) = \nn
&&=(q\bar{q})^{-{1\over 12}+\phi^2}
 \cdot  \prod_{n=1}^{\infty}
(1-q^{n-{1\over 2}+\phi}e^{-2\pi i\theta})^2
(1-q^{n-{1\over 2}-\phi}e^{2\pi i\theta})^2
(1-\bar{q}^{n-{1\over 2}-\phi}e^{-2\pi i\theta})^2
(1-\bar{q}^{n-{1\over 2}+\phi}e^{2\pi i\theta})^2
\nn
&&= {1\over\left|\eta(\tau)\right|^4} \left|\cthe{\phi}{-\theta\!+\!1/2}(0,\tau)\right|^4 . \nn
\label{unorbfer}
\end{eqnarray}
Note that if $\phi = -1/2$, the different $U(1)$-charges of the fermionic zero-modes must be taken into account, but 
the result is the same.

In the twisted sector ($\phi_0=1/2$),  
the case (i) is applicable for $-1/2\le\phi<0$, while the case
(ii) for $0\le\phi < 1/2$.
One can see that for $-1/2\le\phi<0$,
\begin{eqnarray}
&&\trphi \left(q^{L_0}\bar{q}^{\bar{L}_0} e^{-2\pi i\theta (J_0+\bar{J}_0)}\right) 
= \nn
&&=(q\bar{q})^{-{1\over 12}+({1\over 2}+\phi)^2} 
\cdot \prod_{n=1}^{\infty}
(1+q^{n+\phi}e^{-2\pi i\theta})^2(1+q^{n-\phi-1}e^{2\pi i\theta})^2
(1+\bar{q}^{n-\phi-1}e^{-2\pi i\theta})^2(1+\bar{q}^{n+\phi}e^{2\pi i\theta})^2 
\nn
&&= {1\over\left|\eta(\tau)\right|^4}
\left|\cthe{\phi\!+\!1/2}{-\theta}(0,\tau)\right|^4 \ , \nn
\nn
&&\trphi \left( {\bf r} \cdot q^{L_0}\bar{q}^{\bar{L}_0} e^{-2\pi i\theta (J_0+\bar{J}_0)}\right)  
= \nn
&&= (q\bar{q})^{-{1\over 12}+({1\over 2}+\phi)^2}
\cdot  \prod_{n=1}^{\infty}
(1-q^{n+\phi}e^{-2\pi i\theta})^2(1-q^{n-\phi-1}e^{2\pi i\theta})^2
(1-\bar{q}^{n-\phi-1}e^{-2\pi i\theta})^2(1-\bar{q}^{n+\phi}e^{2\pi i\theta})^2 
\nn
&&= {1\over\left|\eta(\tau)\right|^4}
\left|\cthe{\phi\!+\!1/2}{-\theta\!+\!1/2}(0,\tau)\right|^4 \ . \nn
\label {fermionpf}
\end{eqnarray}

When $0\le\phi<1/2$ we obtain
\begin{eqnarray}
&&\trphi \left(q^{L_0}\bar{q}^{\bar{L}_0} e^{-2\pi i\theta (J_0+\bar{J}_0)}\right)
= \nn
&&= (q\bar{q})^{-{1\over 12}+({1\over 2}-\phi)^2} 
\cdot  \prod_{n=1}^{\infty}
(1+q^{n+\phi-1}e^{-2\pi i\theta})^2(1+q^{n-\phi}e^{2\pi i\theta})^2
(1+\bar{q}^{n-\phi}e^{-2\pi i\theta})^2(1+\bar{q}^{n+\phi-1}e^{2\pi i\theta})^2 
\nn
&&= {1\over\left|\eta(\tau)\right|^4} \left|\cthe{\phi\!-\!1/2}{-\theta}(0,\tau)\right|^4  
= {1\over\left|\eta(\tau)\right|^4} \left|\cthe{\phi\!+\!1/2}{-\theta}(0,\tau)\right|^4 \ , \nn
\nn
&&\trphi \left( {\bf r} \cdot q^{L_0}\bar{q}^{\bar{L}_0} e^{-2\pi i\theta (J_0+\bar{J}_0)}\right)
=\nn
&&=(q\bar{q})^{-{1\over 12}+({1\over 2}-\phi)^2}
\cdot  \prod_{n=1}^{\infty}
(1-q^{n+\phi-1}e^{-2\pi i\theta})^2(1-q^{n-\phi}e^{2\pi i\theta})^2
(1-\bar{q}^{n-\phi}e^{-2\pi i\theta})^2(1-\bar{q}^{n+\phi-1}e^{2\pi i\theta})^2 
\nn
&&= {1\over\left|\eta(\tau)\right|^4} \left|\cthe{\phi\!-\!1/2}{-\theta\!+\!1/2}(0,\tau)\right|^4
= {1\over\left|\eta(\tau)\right|^4} \left|\cthe{\phi\!+\!1/2}{-\theta\!+\!1/2}(0,\tau)\right|^4
 \ . \nn
\end{eqnarray}
Thus we get the same result as in (\ref {fermionpf}).

Note that if $\phi = 0$, the different $U(1)$-charges of the fermionic zero-modes must be taken into account, but 
the result is the same.

\subsubsection{Ghosts}

The $N=2$ fermionic ghosts consist of two sets of holomorphic and
antiholomorphic ghosts. One is the usual conformal ghost - anti-ghost pair
denoted as $(b,c)$ with $\lambda=2$.
The other set is the ghost and anti-ghost pair $(\hat{b},\hat{c})$,
which is associated with the $U(1)$ gauge symmetry and has $\lambda=1$.
The corresponding antiholomorphic sets are denoted $(\bar{b},\bar{c})$ and $(\bar{\hat{b}},\bar{\hat{c}})$.
There are also two holomorphic sets of 
superghosts, related to
the two generators of supersymmetry $G^{\pm}$; 
They are denoted $(\beta^{\pm}, \gamma^{\mp})$ (the sign indicates the $U(1)$-charge), and both have $\lambda=3/2$.
The corresponding antiholomorphic sets are denoted $(\bar{\beta}^{\pm}, \bar{\gamma}^{\mp})$.

We shall discuss the holomorphic (anti-)ghosts and (anti-)superghosts. 
A similar discussion applies to 
the antiholomorphic ones, but with $\phi\rightarrow -\phi$ (as in fermions).
A general holomorphic ghost (or superghost) system $(B,C)$ with $h(B)=\lambda,~h(C)=1-\lambda$, obeying 
$BC=-\epsilon CB,~\epsilon=\pm 1$, has the following stress tensor:
\begin{equation}
T(z)=\del B\,C-\lambda\,\del(BC) \ .
\end{equation}
The mode expansions for $(B,C)$ take the form
\begin{equation}
B(z)=\sum_r {B_r\over z^{r+\lambda}} \ ,~~~
C(z)=\sum_s {C_s\over z^{s+1-\lambda}} \ .
\label{bc;1}
\end{equation}
Here
\begin{equation}
r \in \bZ \ ,~~~
s \in \bZ \ ,
\end{equation}
for the (anti-)ghosts ($\epsilon=1$), and:
\begin{equation}
r \in \bZ+{1\over2} \pm \phi \ ,~~~
s \in \bZ+{1\over2} \mp \phi \ ,
\end{equation}
for the (anti-)superghosts ($\epsilon=-1$), where the sign in front of $\phi$ is according to their $U(1)$-charge.

The anti-commutation (for $\epsilon=1$) or commutation (for $\epsilon=-1$) relations are given by
\begin{eqnarray}
\{C_s,B_r\}=\delta_{r+s} &,& [C_s,B_r]=\delta_{r+s}
\end{eqnarray}
The Virasoro generators take the form
\begin{equation}
L_n=\sum_r (n\lambda-r):B_r C_{n-r}:+a\,\delta_n  \ .
\end{equation}
Here $a$ is a zero point energy which depends on the definition of
the ground state, namely a prescription
of the operator ordering $:~:$. 
The ground state $\gs_{\rm gh}$ is defined as
\begin{equation}
B_r \gs_{\rm gh}=0 \ , ~{\rm for}~r\ge 0 \ ,~~~~
C_s \gs_{\rm gh}=0 \ , ~{\rm for}~s> 0 \ .
\label{ghgs}
\end{equation}

The partition function does not include a sum over the fermionic ghosts zero-modes, because of the ghost and anti-ghost insertions 
in the torus amplitude (as in (\ref{measure}) for the $U(1)$ ghosts insertions).

The constant $a$ in the $w$ coordinate reads:

(1) For each set of holomorphic fermionic ghost and anti-ghost $\epsilon=+1$ :
\begin{equation}
a= -\!\!\sum_{n=1,2,\cdots}\!\! n= {1\over 12} \ .
\label{ghostzpe}
\end{equation}

(2) For each set of holomorphic superghost and anti-superghost $\epsilon=-1$ :
\begin{equation}
a= {1\over2} \sum_{n=0,1,\cdots}\!\! \left(n+{1\over 2}+\phi\right) 
+ {1\over2} \sum_{n=0,1,\cdots}\!\! \left(n+{1\over 2}-\phi\right) 
= {1\over 24}-{\phi^2\over 2} \ .
\label{superghostzpe}
\end{equation}

All the (anti-)ghosts and (anti-)superghosts are singlets under the orbifold group, so one obtains:
\begin{eqnarray}
Z_{\rm gh}&\!=\!&
\trphi \left( (-1)^{F_{\rm gh}}b_0 c_0\bar{b}_0\bar{c}_0\,\hat{b}_0\hat{c}_0\bar{\hat{b}}_0\bar{\hat{c}}_0\,
q^{L_0}\bar{q}^{\bar{L}_0} e^{-2\pi i\theta (J_0+\bar{J}_0)}\right)
\nn
&\!=\!&
\left|q^{1\over 6}\prod_{n=1}^{\infty}(1-q^n)^4 \right|^2 \cdot
(q\bar{q})^{{1\over 12}-\phi^2} \nn
&\cdot &\prod_{n=1}^{\infty}
(1+q^{n-{1\over 2}+\phi}e^{-2\pi i\theta})^{-2} (1+q^{n-{1\over 2}-\phi}e^{2\pi i\theta})^{-2} 
(1+\bar{q}^{n-{1\over 2}-\phi}e^{-2\pi i\theta})^{-2} (1+\bar{q}^{n-{1\over 2}+\phi}e^{2\pi i\theta})^{-2} 
\nn
&\!=\!&
\left|\eta(\tau)\right|^{12} \, \left|\cthe{\phi}{-\theta}(0,\tau)\right|^{-4} \ .
\label {ghostpf}
\end{eqnarray}

\vskip 0.3cm
\noindent
{\bf Dependence on the $U(1)$ Charge}

It is easy to see that in all the parts of the partition function
calculated above, the $\phi$ dependence enters into $L_0$, $\tilde{L}_0$
through the form 
$J_0\cdot\phi$ and $-\bar{J}_0\cdot\phi$. 
Instead of tracing over oscillators with different $\phi$'s (and different conformal 
weights), we may sum only over oscillators with $\phi=0$, but add 
$q^{J_0\cdot\phi} \bar{q}^{-\bar{J}_0\cdot\phi}=e^{2\pi i (J_0\tau\phi+\bar{J}_0\bar{\tau}\phi)}$ to the trace,
and integrate over $\phi$. Thus the full dependence on the $U(1)$ charge is 
$e^{2\pi i [J_0(-\theta+\tau\phi)+\bar{J}_0(-\theta+\bar{\tau}\phi)]}= e^{2\pi i (J_0 u + \bar{J}_0 \bar{u})}$. 
This confirms (\ref{u1dep}).

\section{Spectrum of the $N=2$ Orbifold}

\vskip 0.3cm
\noindent
{\bf Different Pictures}

As in the $N=1$ fermionic
string, pictures play an important role in the $N=2$ string.
A picture is defined by the superghost sector ground state \cite{Friedan:1985ge}.
Three pictures will be mostly relevant to us.
\begin{itemize}

\item{} NS boundary conditions ($\phi=0$):
the ground states in the $(-1,-1)$, and $(0,0)$ pictures are defined by
\begin{eqnarray}
\beta^\pm_r \gs_{(-1,-1)}=0 \ , ~{\rm for}~r\ge 1/2   &,& \gamma^\pm_s \gs_{(-1,-1)}=0 \ , ~{\rm for}~s\ge 1/2 \ ,
\label{-1pic}
\end{eqnarray}
\begin{eqnarray}
\beta^\pm_r \gs_{(0,0)}=0 \ , ~{\rm for}~r\ge -1/2   &,& \gamma^\pm_s \gs_{(0,0)}=0 \ , ~{\rm for}~s\ge 3/2 \ .
\label{0pic}
\end{eqnarray}

\item{} R boundary conditions ($\phi=-1/2$):
the ground state in the $(-1/2,-1/2)$ picture is defined by

\begin{eqnarray}
\beta^\pm_r \gs_{(-1/2,-1/2)}=0 \ , ~{\rm for}~r\ge 0   &,& \gamma^\pm_s \gs_{(-1/2,-1/2)}=0 \ , ~{\rm for}~s\ge 1 
\ .
\end{eqnarray}

\end{itemize}

Every state in the $(-1,-1)$ picture has a physically equivalent state in the
$(0,0)$ picture, which is related to it
 by the picture changing operator. Its matter part is 
$G^{M~+}_{-1/2}G^{M~-}_{-1/2}$ \cite{Ooguri:1991fp,Gluck:2003wg}, where $G^{\pm}$
are the two spin $\frac{3}{2}$ generators of the $N=2$ superconformal algebra. 
However, the converse is not true:
not every state in the $(0,0)$ picture has a physically equivalent state 
in the $(-1,-1)$ picture.

\subsection{BRST Analysis: Untwisted Sector}

We will consider only
the massless states, since  there are no massive states 
in the spectrum.
The spectrum of the untwisted sector is found by taking $\bZ_2$ invariant combinations of states of the unorbifolded 
theory. We begin by finding the spectrum of the latter.
We will work in the holomorphic side.
The zero point energy in the $(-1,-1)$, and $(-1/2,-1/2)$ pictures is zero.
Thus, the massless states in these pictures 
are at level zero.

Acting with the BRST charge $Q_{BRST}$ of the $N=2$ strings
(see \cite{Li:1992rr} and the appendix B)
on a level zero state $|\varphi\rangle$, 
produces a state which includes only 
ghosts zero modes $c_0$, $\hat{c}_0$, and 
in addition $\gamma^\pm_0$ in the $(-1/2,-1/2)$ picture.
Also, every physical state is annihilated
by the anti ghosts zero modes $b_0$, $\hat{b}_0$, 
and in addition $\beta^\pm_0$ in the $(-1/2,-1/2)$ picture. 
Therefor, for $|\varphi\rangle$ to 
be BRST invariant it is sufficient 
to require
\begin{eqnarray}
b_0 Q_{BRST}|\varphi\rangle &=& \{b_0, Q_{BRST}\}|\varphi\rangle=L_0|\varphi\rangle = 0 \ , \nn
\hat{b}_0 Q_{BRST}|\varphi\rangle &=& \{\hat{b}_0, Q_{BRST}\}|\varphi\rangle=J_0|\varphi\rangle
= 0  \ , \nn
\beta^\pm_0 Q_{BRST}|\varphi\rangle &=& [\beta^\pm_0, Q_{BRST}]|\varphi\rangle=-G^\pm_0|\varphi\rangle = 0 \ ,
\label{chargevanish}
\end{eqnarray}
with the last equation being relevant only for the $(-1/2,-1/2)$ picture.
In order for
$|\varphi\rangle$ not to be BRST exact, 
it should not
include ghost zero modes $c_0$, $\hat{c}_0$ 
and $\gamma^\pm_0$.
Solving these conditions we get for the BRST cohomology:
\begin{itemize}
\item{} $(-1,-1)$ picture: the state $|0,k\rangle_{(-1,-1)}$ 
with $k^2=0$ is annihilated by $L_0$ and $J_0$.
This is a massless scalar field \cite{Ooguri:1991fp}.

\item{} $(-1/2,-1/2)$ picture: define the ground state by ${\bar \psi}^i_0 |0,k\rangle_{(-1/2,-1/2)} = 0$, $i=1,2$.
Then the state
\begin{equation}
\bar{k}_i\psi^i_0|0,k\rangle_{(-1/2,-1/2)} \ ,
\label{halfscalar}
\end{equation}
with $k^2=0$ is in the BRST cohomology.
We will see later that this state is equivalent to the state
$|0,k\rangle_{(-1,-1)}$.

\item{} $(0,0)$ picture: 
here
\begin{equation}
L_0\gs_{(0,0)} = \left({\alpha^\prime\over 4}k^2 +{1\over 2} (\beta^+_{-1/2}\gamma^-_{1/2}+\beta^-_{-1/2}\gamma^+_{1/2})\right)\gs_{(0,0)} = 
({\alpha^\prime\over 4}k^2-1)\gs_{(0,0)} \ .
\end{equation}
Therefor the massless states are at level one in this picture.
$\gamma^\pm_{1/2}$ are now creation operators, and we can write any charge zero
level one 
state ${\cal O}|0,k\rangle_{(0,0)}$ as
\begin{equation}
{\cal O}|0,k\rangle_{(0,0)} = \sum_n {\cal O}_{n,m} {\gamma^+_{1/2}}^n {\gamma^-_{1/2}}^m |0,k\rangle_{(0,0)} \ ,
\label{O}
\end{equation}
where ${\cal O}_{n,m}$ has charge $m-n$, conformal dimension $1+{1\over 2}(n+m)$, and
does not include $\gamma^\pm_{1/2}$ oscillators.
As we will show, we find the following physical states:
\vskip 0.2cm
(1) $\left(\bar{k}_i\alpha^i_{-1} - k_i\bar{\alpha}^i_{-1} - (\bar{k}_i\psi^i_{-1/2})(k_i\bar{\psi}^i_{-1/2})\right)
|0,k\rangle_{(0,0)}$, plus terms with ghosts oscillators.\\
This is the massless scalar in the $(0,0)$ picture.\\
\vskip 0.1cm
(2)$\left(\alpha^i_{-1}-\sqrt{2}\gamma^-_{1/2}b_{-1}\psi^i_{-1/2}\right)|0,0\rangle_{(0,0)}$,
$\left(\bar{\alpha}^i_{-1}-\sqrt{2}\gamma^+_{1/2}b_{-1}\bar{\psi}^i_{-1/2}\right)|0,0\rangle_{(0,0)}$.
\footnote{For an earlier work on discrete states in $N=2$ strings on a
flat background, see \cite{Bienkowska:1991zs}.}
Combined with their anti holomorphic parts,
these are global dilaton, B-field and metric degrees of freedom.

\end{itemize}

Let us see that. First, note that for any physical state of the form (\ref{O}),
when it is inserted in a scattering amplitude or an inner product, 
only 
${\cal O}_{0,0}$ contributes.
This is since the $\gamma^\pm_{1/2}$ oscillators annihilate the other parts of the 
scattering amplitude or inner product. 
For instance, the inner product of $\gamma^\pm_{1/2}|0,k\rangle_{(0,0)}$ with itself 
is zero.

If ${\cal O}|0,k\rangle_{(0,0)}$ is BRST invariant, it obeys 
\begin{eqnarray}
G^\pm_{-1/2}{\cal O}|0,k\rangle_{(0,0)} &=& -[\beta^\pm_{-1/2},Q_{BRST}]{\cal O}|0,k\rangle_{(0,0)} \nn
&=& Q_{BRST}\beta^\pm_{-1/2}{\cal O}|0,k\rangle_{(0,0)} = -Q_{BRST}{\cal O}_{\pm 1}|0,k\rangle_{(0,0)} + ... \nn
&=& -\left(L^M_{-1}c_1 + J^M_{-1}\hat{c}_1 + G^{M\pm}_{-3/2}\gamma^\mp_{3/2}\right){\cal O}_{\pm 1}|0,k\rangle_{(0,0)} + 
...\nn
&=& \left(L^M_{-1}A^\pm + J^M_{-1}B^\pm + G^{M\pm}_{-3/2}C \right)|0,k\rangle_{(0,0)} + ... \ ,
\end{eqnarray}
where dots refer to terms with ghost oscillators. 
$A^\pm$ and $B^\pm$ are operators of charge $\pm 1$ and 
conformal dimension $1/2$, and independent of ghost oscillators.
$C$ is a number. ${\cal O}_{\pm 1}$ is a short notation for ${\cal O}_{0,1}$ and ${\cal O}_{1,0}$.
The matter part of $G^\pm_{-1/2}{\cal O}|0,k\rangle_{(0,0)}$ is $G^{M\pm}_{-1/2}{\cal O}^M|0,k\rangle_{(0,0)}$ and thus
\begin{equation}
G^{M\pm}_{-1/2}{\cal O}^M|0,k\rangle_{(0,0)}
= \left(L^M_{-1}A^\pm + J^M_{-1}B^\pm + G^{M\pm}_{-3/2}C \right)|0,k\rangle_{(0,0)} \ .
\label{OCQG}
\end{equation}

${\cal O}^M|0,k\rangle_{(0,0)}$ is a level one state with charge zero, and no ghost oscillators. 
Therefor it is of the 
form
\begin{equation}
{\cal O}^M|0,k\rangle_{(0,0)} = \left(A_i\alpha^i_{-1} + A^\prime_i\bar{\alpha}^i_{-1} + 
P_{ij}\psi^i_{-1/2}\bar{\psi}^j_{-1/2} \right) |0,k\rangle_{(0,0)} \ .
\end{equation}
For $k\ne0$, any solution of (\ref{OCQG}) 
is a linear combination of state (1) and the null states\\
${1\over 2}(\bar{k}_i\alpha^i_{-1} + k_i\bar{\alpha}^i_{-1})|0,k\rangle_{(0,0)} = L^M_{-1}|0,k\rangle_{(0,0)}$,
which is the matter part of $Q_{BRST}b_{-1}|0,k\rangle_{(0,0)}$, and 
${1\over 2}\psi_{i~-1/2}\bar{\psi}^i_{-1/2}|0,k\rangle_{(0,0)} = J^M_{-1}|0,k\rangle_{(0,0)}$,
which is the matter part of $Q_{BRST}\hat{b}_{-1}|0,k\rangle_{(0,0)}$.
For $k=0$, the solutions of (\ref{OCQG}) 
are linear combinations of (2) and the null state $J^M_{-1}|0,0\rangle_{(0,0)}$.

The untwisted sector of the theory on the
orbifold consists of $\bZ_2$-invariant combinations of states (1) and (2). Therefor 
the massless scalars appear in combinations of the form $(|0,k\rangle+|0,-k\rangle)$.


\subsection{BRST Analysis: Twisted Sector}

The boundary conditions for the matter fields imply that $\partial X^i$, 
$\partial\bar{X^i}$ are expanded in half integer oscillators $\alpha^i_{n+1/2}$, $\bar{\alpha}^i_{n+1/2}$,
and therefor there are no momentum degrees of freedom.
$\psi^i$ and ${\bar \psi}^i$ are expanded in half 
integer oscillators in the $(-1/2,-1/2)$ picture, and in integer ones in 
the $(-1,-1)$ and $(0,0)$ pictures. 
The same holds for the anti holomorphic side.
$L_0$ and $J_0$ 
have different normal ordering constants relative to the untwisted sector. 

\begin{itemize}
\item{} $(-1,-1)$ picture:
the normal ordering constant of $L_0$ is $+1/2$,
and condition (\ref{chargevanish}) cannot be satisfied.

\item{} $(-1/2,-1/2)$ picture: $L_0$ has zero normal ordering constant. 
The ground state  is annihilated by $J_0$ and $G^\pm_0$ 
and is the only physical 
state.

\item{} $(0,0)$ picture: 
the normal ordering constants of $L_0$ and  $J_0$
are $-1/2$ and $-1$, respectively. Therefor the matter 
part of any physical state takes the form
\begin{equation}
\left( C_{ij}\alpha^i_{-1/2}\psi^j_0 + D_{ij}\bar{\alpha}^i_{-1/2}\psi^j_0 \right) |\sigma\rangle_{(0,0)} \ ,
\end{equation}
where the  
twisted sector ground state $|\sigma\rangle_{(0,0)}$ is annihilated by $\bar{\psi}^i_0$,
$i=1,2$. 
Following the steps leading to (\ref{OCQG}), 
we get
\begin{eqnarray}
G^{M\pm}_{-1/2}{\cal O}^M|\sigma\rangle_{(0,0)} &=& 
\left(L^M_{-1}A^\pm + J^M_{-1}B^\pm\right)|\sigma\rangle_{(0,0)} \ ,
\end{eqnarray}
where $A^\pm$, $B^\pm$ have conformal dimension zero and charge two
 (for the plus sign), and zero (for the minus sign).
Therefor $C_{ij}$ is proportional to $\epsilon_{ij}$ and $D_{ij}$ is proportional to $\eta_{ij}$,
and we get two physical states, whose matter parts read 
\begin{eqnarray}
\sqrt{2}\epsilon_{ij}\alpha^i_{-1/2}\psi^j_0 |\sigma\rangle_{(0,0)} 
&=& {1\over \sqrt{2}} \alpha^i_{-1/2}\bar{\psi}_{i~0} \psi^1_0 \psi^2_0 |\sigma\rangle_{(0,0)} 
= G^{-~M}_{-1/2} \psi^1_0 \psi^2_0 |\sigma\rangle_{(0,0)} \ , \nn
{1\over \sqrt{2}} \eta_{ij}\bar{\alpha}^i_{-1/2}\psi^j_0 |\sigma\rangle_{(0,0)} 
&=& G^{+~M}_{-1/2} |\sigma\rangle_{(0,0)} \ .
\end{eqnarray}
These states have corresponding states in the $(-1,0)$ picture 
and in the $(0,-1)$ picture\footnote{The ground states in these pictures satisfy (\ref{-1pic}) for one 
$(\beta,\gamma)$ system, and (\ref{0pic}) for the other $(\beta,\gamma)$ system.}
\begin{eqnarray}
\psi^1_0 \psi^2_0 |\sigma\rangle_{(-1,0)} &,& |\sigma\rangle_{(0,-1)} \ .
\label{01ts}
\end{eqnarray}
We will later see that 
these state are physically equivalent to the twisted ground state in the $(-1/2,-1/2)$ picture, and  
there is only one twisted state.

\end{itemize}

\subsection{Vertex Operators}

Denote by $\sigma^i$ and $s^i$,  $i=1,2$ , the bosonic and fermionic
twist fields for the  $\bZ_2$ orbifold \cite{Dixon:1987}.
Their OPEs (omitting index $i$) read
\begin{eqnarray}
\partial X(z)\sigma(w,\bar{w})\sim (z-w)^{-1/2}\tau(w,\bar{w}) &,&
\psi(z)s(w)\sim (z-w)^{-1/2}t(w) \ , \nn
\partial \bar{X}(z)\sigma(w,\bar{w})\sim (z-w)^{-1/2}\tau^{\prime}(w,\bar{w}) &,& 
\bar{\psi}(z)s(w)\sim (z-w)^{1/2}t^{\prime}(w) \ , 
\label{clOPE}
\end{eqnarray}
where  $\tau$, $\tau^{\prime}$ and  $t$, $t^{\prime}$ are
excited bosonic and fermionic twist fields, respectively.
Bosonizing the fermionic fields
\begin{eqnarray}
\psi^i = e^{iH_i} \ , ~~~
\bar{\psi_i} = e^{-iH_i} \ &,& H_i(z)H_i(w) \sim  -\log (z-w) \ ,
\end{eqnarray}
We obtain
$s^i = e^{-iH_i/2}$.
Consider next the construction of the vertex operators.
\vskip 0.3cm
\noindent
{\bf untwisted sector}

The ghost part of the vertex operator corresponding to a state in the $(n,m)$ picture is
\begin{equation}
V_g^{(n,m)} =
c \, e^{n\phi^+} \, e^{m\phi^-} \, (z)  \, \bar{c} \, e^{n{\bar \phi}^+} \, e^{m{\bar \phi}^-} \, (\bar{z}) \ ,
\label {nmpic}
\end{equation}
where $\beta^{\mp}\gamma^{\pm} = \partial \phi^{\pm}$.

\begin{itemize}

\item{} $(-1,-1)$ picture:
the  vertex operator corresponding to the
 scalar takes the form
\begin{equation}
V_g^{(-1,-1)}\cdot e^{i(k_i\bar{X}^i(z,\bar{z})+\bar{k}_iX^i(z,\bar{z}))} \ .
\end{equation}

\item{} $(-1/2,-1/2)$ picture:
the  vertex operator corresponding to the
scalar (\ref{halfscalar}) takes the form
\begin{equation}
V_g^{(-1/2,-1/2)}\cdot \, (\bar{k}_i \psi^i \, e^{-iH_1/2} e^{-iH_2/2} \, (z)\cdot (h.c))  
\cdot e^{i(k_i\bar{X}^i(z,\bar{z})+\bar{k}_iX^i(z,\bar{z}))} \ .
\label{halfpicu}
\end{equation}

\item{} $(-1,0)$ picture:
The matter part of the scalar state in this picture is $G^{M+}_{-1/2}|0,k\rangle_{(-1,0)}$
and the corresponding vertex operator 
takes the form
\begin{equation}
V_g^{(-1,0)}\cdot \, (\bar{k}_i \psi^i \, (z)\cdot (h.c))  
\cdot e^{i(k_i\bar{X}^i(z,\bar{z})+\bar{k}_iX^i(z,\bar{z}))} \ .
\label{halfpicu2}
\end{equation}

\item{} $(0,0)$ picture:
apart from the one corresponding to the massless scalar,
we found additional states 
in the $(0,0)$ picture compared
to the $(-1,-1)$ picture.
These are global degrees of freedom of the form
\begin{equation}
c\bar{c}\cdot \left(G_{ij} \partial X^i \bar{\partial} \bar{X}^j(z,\bar{z}) 
+ G^\prime_{ij} \partial X^i \bar{\partial} X^j(z,\bar{z}) + 
\widetilde{G}_{ij} \partial \bar{X}^i \bar{\partial} X^j(z,\bar{z}) 
+ \widetilde{G}^\prime_{ij}\partial \bar{X}^i \bar{\partial} \bar{X}^j(z,\bar{z}) \right) \ .
\end{equation}
$G$, $G^\prime$, $\widetilde{G}$ and $\widetilde{G}^\prime$ are number matrices.

\end{itemize}
\vskip 0.3cm
\noindent
{\bf Twisted sector}

\begin{itemize}
\item{} $(-1/2,-1/2)$ picture:
the twisted state is represented by the vertex operator
\begin{equation}
V_g^{(-1/2,-1/2)}\cdot \, \sigma^1(z,\bar{z})\,\sigma^2(z,\bar{z}) \ .
\label{halfpict}
\end{equation}

\item{} $(-1,0)$ picture:
the twisted state (\ref{01ts}) is represented by the vertex operator
\beq
V_g^{(-1,0)}\cdot\, e^{iH_1/2} \, e^{iH_2/2} (z) \cdot (h.c.) \cdot \, \sigma^1(z,\bar{z})\,\sigma^2(z,\bar{z}) 
\ .
\label{01pict}
\eeq

\item{} $(0,-1)$ picture:
the twisted state (\ref{01ts}) is represented by the vertex operator
\beq
V_g^{(0,-1)} \cdot
\, e^{-iH_1/2} \, e^{-iH_2/2} (z) \cdot (h.c.) \cdot \, \sigma^1(z,\bar{z})\,\sigma^2(z,\bar{z}) \ .
\label{01pict1}
\eeq

\end{itemize}


\vskip 0.3cm
\noindent
{\bf Spectral Flow}

Spectral flows \cite{Schwimmer:1986mf} are automorphisms of the
 $N=2$ SCA. They are characterized by a single parameter
$\eta$ and map the $N=2$ generators as 
\begin{eqnarray}
L_n^{\eta}&\!=\!& L_n+\eta J_n+{c\over 6}\eta^2\,\delta_n   \ , \nn
J_n^{\eta}&\!=\!& J_n+{c\over 3}\eta\,\delta_n   \ , \nn
G_r^{\eta \pm}&\!=\!& G^{\pm}_{r\pm\eta} \ .  
\label{spectralflow}
\end{eqnarray}
A spectral flow operator which implements these transformations 
takes the form
\begin{equation}
U_{\eta}=
e^{-\eta(iH_1 + iH_2 + \phi^- - \phi^+ - \hat b c)} = e^{-\eta(iH_1 + iH_2 + \phi^- - \phi^+)}(1+\eta\hat b c) \ .
\end{equation}
Note that the $U(1)$ current takes the form
\begin{equation}
J = \partial(iH_1 + iH_2 + \phi^- - \phi^+ - \hat b c) = \partial \Phi \ ,   
\end{equation}
and $U_{\eta} = e^{-\eta\Phi}$.

$U_{\eta}$ 
maps a state in the $(n,m)$-picture (\ref{nmpic}) to one in the $(n+\eta,m-\eta)$-picture.
and $\eta$ can be regarded as a Wilson line, turned on 
around the state.
In particular,
the spectral flows can map a state
in the $(-1/2,-1/2)$ picture to a state in the $(-1,0)$ and $(0,-1)$ pictures, but not
to a state in the  $(-1,-1)$. For example, the state
(\ref{halfpict}) gets mapped to (\ref{01pict}) and (\ref{01pict1})
by the spectral flow with $\eta = \pm 1/2$, 
and 
the state (\ref{halfpicu}) to (\ref{halfpicu2}) by the spectral flow with $\eta = -1/2$.

\vskip 0.3cm
\noindent
{\bf Summary}

The untwisted sector consists of $\bZ_2$ invariant combinations of the states in the unorbifolded 
theory. These are a massless scalar, and global degrees of freedom which
describe the dilaton, metric and two-form field.
The twisted sector consists of one twisted state, 
described, for instance,  by the twisted ground state in the $(-1/2,-1/2)$ picture.

\section{Four Twisted States Scattering Amplitude}

\subsection{Finiteness and Invariance Under Crossing Symmetry}

By using
\beq
x=\theta_2^4(\tau)/\theta_3^4(\tau),~~~~F(x)=\theta_3^2(\tau),~~~~
1-x=\theta_4^4(\tau)/\theta_3^4(\tau) \ .
\eeq
one finds that
\begin{eqnarray}
{\cal A}_{tttt} 
&=& \int dx d\bar{x} \, {|x(1-x)|^{-4/3}{|\theta_2(\tau)\theta_3(\tau)\theta_4(\tau)|}^{-8/3}\over \tau_2^2(x,\bar{x})} 
V_{\Lambda} \sum_{v_1\in\Lambda_c,v_2\in\Lambda_{c'}}e^{-{\pi\over 4\tau_2}[v_2\bar{v}_2-\tau_1(v_1\bar{v}_2+\bar{v}_1v_2)+
|\tau|^2v_1\bar{v}_1]} \nn
&=& \int dx d\bar{x} \, {|4x(1-x)|^{-4/3}\over |\eta(\tau)|^8\tau_2^2(x,\bar{x})} 
V_{\Lambda} \sum_{v_1\in\Lambda_c,v_2\in\Lambda_{c'}}e^{-{\pi\over 4\tau_2}[v_2\bar{v}_2-\tau_1(v_1\bar{v}_2+\bar{v}_1v_2)+
|\tau|^2v_1\bar{v}_1]} \ .
\label{crossing}
\end{eqnarray}
With this, crossing symmetry is manifest:
The transformations $\tau\rightarrow -1/\tau$ and
$\tau\rightarrow\tau+1$ amount to 
$x\rightarrow 1-x$ and $x\rightarrow x/(x-1)$, respectively.
%

We will now show that (\ref{4ptfun_comp}) is finite. Because of crossing symmetry, 
it is enough to show that the integral converges 
at $x\rightarrow 0$. In this limit $F(x)\rightarrow 1$ and $\tau(x)\sim -{i\over\pi}\ln({x\over 16})$. 
For simplicity we assume $\epsilon_2=0$ and that both complex dimensions of the $T^4$ have the same radius $R$, 
with the sum over $v^i_{1,2}$, $\bar{v}^i_{1,2}$ ($i=1,2$) replaced by a sum over the integers or half-integers 
$n^\mu_{1,2}$ ($\mu=1...4$), so that $v^i_{1,2} = 2R(n^1_{1,2}+ i n^2_{1,2},n^3_{1,2}+i n^4_{1,2})$, 
$n^\mu_1\in\bZ+{1\over 2}\epsilon_1^\mu$ and $n^\mu_2\in\bZ+{1\over 2}\epsilon_3^\mu$.

Taking $x=re^{i\varphi}$, the integral in (\ref{4ptfun_comp}) over a small disk of radius $r_0 <1$ around 
$x=0$ is
\begin{eqnarray}
&&\pi^2 \int_0^{2\pi} d\varphi \, \int_0^{r_0} dr \, {1\over r\ln^2(r/16)}
\sum_{n^\mu_{1,2}\in\bZ+{1\over 2}\epsilon_{1,3}^\mu}
e^{{(\pi R)^2 \over \ln(r/16)}(n_2^\mu - n_1^\mu\varphi/\pi)^2 + R^2\ln(r/16) (n_1^\mu)^2} \nn
&=& \pi^2 \int_0^{2\pi} d\varphi \, \int_0^{y_0} dy \,
\sum_{n^\mu_{1,2}\in\bZ+{1\over 2}\epsilon_{1,3}^\mu}
e^{-(\pi R)^2 y  (n_2^\mu - n_1^\mu\varphi/\pi)^2 - {R^2\over y}(n_1^\mu)^2} \nn
&<& \pi^2 \sum_{n_1^\mu\in\bZ+{1\over 2}\epsilon_1^\mu} e^{-{R^2\over y_0}(n_1^\mu)^2} \int_0^{2\pi} d\varphi 
\sum_{n_2^{\mu~\prime}\in\bZ +{1\over 2}\epsilon_3^\mu - n_1^\mu\varphi/\pi} \int_0^{y_0} dy \, e^{-(\pi R)^2 y  
(n_2^{\mu~\prime})^2} \nn
&\le& 4 \pi^2 \sum_{n_1^\mu\in\bZ} e^{-{R^2\over y_0}(n_1^\mu)^2} \int_0^{2\pi} d\varphi 
\sum_{n_2^{\mu~\prime}\in\bZ} \int_0^{y_0} dy \, e^{-(\pi R)^2 y  (n_2^{\mu~\prime})^2} \nn
&=& 8\pi^3 \sum_{n_1^\mu\in\bZ} e^{-{R^2\over y_0}(n_1^\mu)^2} 
\left(y_0 + \sum_{n_2^{\mu~\prime}\in\bZ-\{0\}} 
{1-e^{-(\pi R)^2 y_0 (n_2^{\mu~\prime})^2}\over (\pi R)^2 (n_2^{\mu~\prime})^2} \right)\nn
&<& {8\pi\over R^2} \sum_{n_1^\mu\in\bZ} e^{-{R^2\over y_0}(n_1^\mu)^2} 
\cdot \left(\pi^2 R^2 y_0+ \sum_{n_2^{\mu~\prime}\in\bZ-\{0\}} {1\over (n_2^{\mu~\prime})^2}\right) = ~finite \ .
\end{eqnarray}
We substituted $y=-{1\over \ln(r/16)}$, and defined $y_0 \equiv-{1\over \ln(r_0/16)}$.
In the third line we defined $n_2^{\mu~\prime} \equiv n_2^\mu - n_1^\mu\varphi/\pi$, and in line four we replaced 
$n_1^\mu$, $n_2^{\mu~\prime}$ by their integer values.

Note that the result is finite for a $(4,0)$ signature. For a $(2,2)$ signature there will be 
infinite values of $n_1^\mu$ and $n_2^{\mu~\prime}$ for which their square is zero. However, the divergence 
of the amplitude in the $(2,2)$ signature is the usual divergence of the path integral in Minkowski 
space, and the amplitude should be calculated by wick rotating it to the $(4,0)$ signature.

${\cal A}_{tttt}$ is finite in the non-compact orbifold $\bC^2/\bZ_2$ as well. 
In the limit $x\rightarrow 0$, $F(x)\rightarrow 1$ and $\tau_2\sim -{1\over\pi}\ln({|x|\over 16})$. Therefor
in (\ref{4ptfun_nc}) the leading singularity of the integrand around $x=0$ is 
${\pi^2\over |x|^2{\ln}^2(|x|/16)}$. The integral converges in this area, since 
${dx d\bar{x}/{(|x|\ln|x|)}^2} \sim dr/r\ln^2r = d(-1/\ln r)$, where $r\equiv |x|$. 
Using crossing symmetry for exploring 
the other non-trivial limits of this integrand, we see that (\ref{4ptfun_nc}) totally converges.

\subsection{Poles}

We shall see which poles in $p_{L,R}$ exist in the amplitude ${\cal A}_{tttt}$ in its form (\ref{Poisson}).

In the limit $x\rightarrow 0$,
\begin{eqnarray}
F(x) = 1+{1\over4}x+... &,& F(1-x) = -{1\over\pi} \left(\ln(x/16) + {1\over 4}x[\ln(x/16)+2] +...\right) \ .
\end{eqnarray}
In particular $w(x)\sim x / 16$ as  $x\rightarrow 0$.
Every term of the sum in the integrand of (\ref{Poisson}) can be expanded around $x=0$ in powers of $x$ and $\bar{x}$, 
so that the contribution of a small disk of radius $a$, $D(a)$, around $x=0$ is, up to the 
$e^{-2\pi i (f_{\epsilon_2}-f_{\epsilon_3})\cdot p}$ phase
\begin{eqnarray}
\int_{D(a)} & dx d\bar{x} & x^{-1+(p+v/2)^2/2} \, \bar{x}^{-1+(p-v/2)^2/2} \cdot 
\sum_{n=0}^\infty A_n x^n  \cdot \sum_{k=0}^\infty \bar{A}_k \bar{x}^k  \nn
&=& 2\pi \int_0^a dr r^{-1+p^2+v^2/4} \cdot r^{n+k} \sum_{n,k=0}^\infty A_n \bar{A}_k \delta_{n-k+p\cdot v} \nn 
&=& 2\pi \sum_{n=0}^\infty {a^{(p+v/2)^2 + 2n}\over (p+v/2)^2 + 2n} A_n \bar{A}_{n+p\cdot v} 
= 2\pi \sum_{k=0}^\infty {a^{(p-v/2)^2 + 2k}\over (p-v/2)^2 + 2k} A_{k-p\cdot v} \bar{A}_k \ .
\label{4tpoles}
\end{eqnarray}
$A_n$, $\bar{A}_k$ are constants which depend only on $(p+v/2)^2$ and $(p-v/2)^2$, respectively. 
In the second line of (\ref{4tpoles}) we changed variables to polar coordinates, and the $\delta$-Kronecker
arises from 
integrating over the angular coordinate.

Equation (\ref{4tpoles}) is  a sum over poles of the form $(p+v/2)^2 + 2n$ (or $(p-v/2)^2 + 2k$). 
If for every $n>0$ 
$~A_n$ is proportional to $(p+v/2)^2 + 2n$, which also means that 
for every $k>0$ $~\bar{A}_k$ is proportional to 
$(p-v/2)^2 + 2k$, then the residues of the poles with $n\ne 0$ or $k\ne 0$ vanish. 
In order to see that this is indeed the case, it is sufficient
 to check that (\ref{4tpoles}) has no such poles in the case 
$v=0$. We checked this explicitly for $0 < n \le 20$ 
\footnote{For example, for $n=1,2$ 
we get $A_1=2^{-2(p+v/2)^2-2}((p+v/2)^2+2)$ and
\\
$A_2=2^{-2(p+v/2)^2-7}(4(p+v/2)^2+13)\cdot((p+v/2)^2+4)$.}.

If (\ref{4tpoles}) has poles only at $n=k=0$, then these poles are at 
$(p + v/2)^2 = (p - v/2)^2 = 0$. Equivalently, the poles are
at $p_L^2 = p_R^2 = 0$.
We find that $A_0 = \bar{A}_0 = 4^{-(p+v/2)^2}$.
Equation (\ref{4tpoles}) takes the form
\begin{eqnarray}
2\pi \, {(a/16)^{p_L^2}\over p_L^2} \, \delta_{p_L^2-p_R^2} + f(a,p_L^2,p_R^2)
\end{eqnarray}
where $f$ has no poles in $p_L^2$, $p_R^2$.

\section{Two Twisted and Two Untwisted States Amplitude}

We start with 
\begin{eqnarray}
{\cal A}_{tt,p,k}&=&\int dzd\bar{z} \, \langle \sigma_1(z_\infty)\sigma_2(z_\infty) \,
{1\over 2}\left(e^{ik_{L,R}\cdot X^{L,R} (w,\bar{w})} + e^{-ik_{L,R}\cdot X^{L,R} (w,\bar{w})}\right) \, \nn
&& {1\over 2}\left(V^{(0,0)}(p_{L,R};z,\bar{z})+V^{(0,0)}(-p_{L,R};z,\bar{z})\right) \, \sigma_1(0)\sigma_2(0)\rangle \nn
&&\cdot|\langle e^{-\phi^+(z_\infty)/2}e^{-\phi^+(w)}e^{-\phi^+(0)/2}\rangle \,
\langle e^{-\phi^-(z_\infty)/2}e^{-\phi^-(w)}e^{-\phi^-(0)/2}\rangle \, \langle c(z_\infty)c(w)c(0)\rangle|^2  \ ,
\nn
\end{eqnarray}
where $k_{L,R}\cdot X^{L,R} \equiv k_L \cdot X^L + k_R \cdot X^R$.

Taking $z_\infty\rightarrow\infty$, the ghost part is $|z_\infty|$, canceling 
the matter part at this limit.
It is easy to see that the fermionic fields in $V^{(0,0)}$ do not contribute to the amplitude, 
and so we are left with:
\begin{eqnarray}
{1\over 4}\int &dzd\bar{z}& \, |z_\infty| \langle \sigma_1(z_\infty)\sigma_2(z_\infty)
\left(e^{ik_{L,R}\cdot X^{L,R} (w,\bar{w})} + e^{-ik_{L,R}\cdot X^{L,R} (w,\bar{w})}\right)
\left(p_L \cJ\partial X\right) \left(p_R \cJ\bar{\partial}X\right) \nn
&\cdot& \left(e^{ip_{L,R}\cdot X^{L,R} (z,\bar{z})} + e^{-ip_{L,R}\cdot X^{L,R} (z,\bar{z})}\right) 
\sigma_1(0)\sigma_2(0)\rangle \ .
\end{eqnarray}

We need to calculate the bosonic fields Green function in the presence of two twists. 
In order to simplify the  notation, 
we will work in one complex dimension.
The Green function is defined as
\begin{eqnarray}
G(z,\bar{z},w,\bar{w}) &\equiv& \langle \sigma(\infty) \, X(z,\bar{z})
 \, \bar{X}(w,\bar{w}) \, \sigma(0)\rangle \nn
&=& -2\left(\int dz dw \, g(z,w) + \int d\bar{z} dw \, h(\bar{z},w) + 
\int dz d\bar{w} \, \bar{h}(z,\bar{w}) + \int d\bar{z} d\bar{w} \, \bar{g}(\bar{z},\bar{w})\right) , \nn
g(z,w) &\equiv& -{1\over 2}\langle \sigma(\infty) \, \partial_z X(z) 
\, \partial_w \bar{X}(w) \, \sigma(0)\rangle \ , \nn
h(\bar{z},w) &\equiv& 
-{1\over 2}\langle \sigma(\infty) \, \bar{\partial}_{\bar{z}}X \, (\bar{z})\partial_w\bar{X}(w)
 \, \sigma(0)\rangle  \ .
\end{eqnarray}
Note that  $\langle\sigma(\infty)\, \sigma(0)\rangle \equiv \lim_{z_{\infty}\rightarrow\infty}
|z_{\infty}|^{1/2}\langle\sigma(z_{\infty})\sigma(0)\rangle = 1$.

Define
\begin{eqnarray}
g(z,w;z_i,\bar{z}_i) &\equiv& -{1\over 2}\langle \sigma(z_1,\bar{z_1}) \, \partial X(z) \, \partial \bar{X}(w) \, 
\sigma(z_2,\bar{z_2})\rangle /\langle \sigma(z_1,\bar{z_1})\sigma(z_2,\bar{z_2})\rangle \ ,
\nn
h(\bar{z},w;z_i,\bar{z}_i) &\equiv& -{1\over 2}\langle \sigma(z_1,\bar{z_1}) \, \bar{\partial} X(\bar{z})
\, \partial \bar{X}(w) \, \sigma(z_2,\bar{z_2})\rangle/\langle \sigma(z_1,\bar{z_1})\sigma(z_2,\bar{z_2})\rangle \ ,
\end{eqnarray}
and $g(z,w) = g(z,w;\infty,0)$, $h(\bar{z},w) = h(\bar{z},w;\infty,0)$.
Using the behavior of $\partial X$ as $z,w\rightarrow z_i$, and the fact
that $g$ has a double pole with residue one
 as 
$z\rightarrow w$ (and  no single pole), while $h$ has no such poles, 
and using the transformations of $g$, $h$ under the coordinate transformations $(z\rightarrow z+a)$, $(z\rightarrow \lambda\cdot z)$,
we find that
\begin{eqnarray}
g(z,w;z_i,\bar{z}_i) &=& (z-z_1)^{-1/2}(z-z_2)^{-1/2}(w-z_1)^{-1/2}(w-z_2)^{-1/2} \nn
&&\cdot\left[{(z-z_1)(w-z_2)+(z-z_2)(w-z_1)\over 2(z-w)^2} + A(z_i,\bar{z}_i)\right] \ , \nn
h(\bar{z},w;z_i,\bar{z}_i) &=& (\bar{z}-\bar{z}_1)^{-1/2}(\bar{z}-\bar{z}_2)^{-1/2}(w-z_1)^{-1/2}(w-z_2)^{-1/2}
B(z_i,\bar{z}_i)\ .
\end{eqnarray}

By using the transformations of $g$, $h$ under the coordinate transformation $(z\rightarrow -1/z)$, 
we see that $A=B=0$.
Setting $z_2=0$ and $z_1\rightarrow\infty$, we have
\begin{eqnarray}
g(z,w) &=& z^{-1/2}w^{-1/2}{z+w\over 2(z-w)^2} \ ,  \nn
h(\bar{z},w) &=& 0
\end{eqnarray}
So we arrive at
\bea
G(z,\bar{z},w,\bar{w}) = -2\left(\int dz dw \, g(z,w) + 
\int d\bar{z} d\bar{w} \, \bar{g}(\bar{z},\bar{w})\right) = 
-2\ln \left|{w^{1/2}-z^{1/2}\over w^{1/2}+z^{1/2}}\right|^2 + C \ .
\eea

The constant $C$ is zero. This can be seen by
computing $G(z,\bar{z},w,\bar{w})$ using the state-operator 
correspondence
\begin{eqnarray}
G(z,\bar{z},w,\bar{w}) &=& \langle \sigma | \, X(z,\bar{z}) \, \bar{X}(w,\bar{w}) \, |\sigma\rangle= 
-\sum_{m,n\in\bZ} \, \langle \sigma | \, \alpha_{m-1/2}{z^{-m+1/2}\over -m+1/2} \, 
\bar{\alpha}_{n-1/2}{w^{-n+1/2}\over -n+1/2} \, |\sigma \rangle \, + \,  \left(c.c\right)\nn
&=& 2\sum_{m>0}{z^{-m+1/2}w^{m-1/2}\over m-1/2} \, + \, \left(c.c\right)
= -2\ln\left|{w^{1/2}-z^{1/2}\over w^{1/2}+z^{1/2}}\right|^2 \ .
\nn
\label{finalG}
\end{eqnarray}

Expression (\ref{finalG}) splits into the holomorphic Green function $X_L X_L$ and the antiholomorphic 
Green function $X_R X_R$. We denote them by $G(z,w)$ and $\bar{G}(\bar{z},\bar{w})$, respectively.
The matter part of the two twisted  two untwisted states
scattering amplitude can then be calculated by the contractions of the 
bosonic matter fields using $G(z,w)$ and $\bar{G}(\bar{z},\bar{w})$ \cite{Dixon:1987}.
Recalling that we have two complex 
dimensions, we get 
\begin{equation}
(z_\infty)^{1/2} \, \langle \sigma_1(z_\infty)\sigma_2(z_\infty) \, 
e^{ik_L\cdot X^L (w)} \, e^{ip_L X^L(z)} \, \sigma_1(0)\sigma_2(0)\rangle
= e^{-k_L\cdot p_L G(z,w)} = \left({w^{1/2}-z^{1/2}\over w^{1/2}+z^{1/2}}\right)^{2k_L\cdot p_L} \ ,
\label{OPEuutt}
\end{equation}
and similarly for the antiholomorphic side.

Every insertion of $\partial X_L^i(z)$, 
$\partial \bar{X}_L^i(z)$, $\bar{\partial} X_R^i(\bar{z})$ or $\bar{\partial} \bar{X}_R^i(\bar{z})$ 
results in multiplying 
the result of (\ref{OPEuutt}) by 
\beq
ik_L^i\partial_z G(z,w) = -2ik_L^i w^{1/2}z^{-1/2} / (z-w) \ ,
\eeq
and its corresponding 
conjugates, respectively (due to its contraction with $e^{ik_{L,R}\cdot X^{L,R} (w)}$).

Note that the on-shell conditions $p_L^2=k_L^2=0$ implies that
$2k_L\cdot p_L = (k_L+p_L)^2 = -(k_L-p_L)^2$. 
Therefor the r.h.s of (\ref{OPEuutt}) diverges in the limit $z\rightarrow (1\pm i\epsilon) w$ for $(k_L\pm p_L)^2 <0$.
It is instructive to
compare this with the unorbifolded OPE
\beq
 e^{ik_L X^L(w)} e^{ip_L X^L(z)} \sim 
(w-z)^{(k_L+p_L)^2/2}e^{i(k_L+p_L)X^L} \ ,
\eeq
which diverges for $(k_L+p_L)^2<0$ in the limit $z\rightarrow w$,
and represents the appearance of an intermediate state of (left) momentum $p_L+k_L$. 
On the orbifold 
a similar thing happens, but two intermediate states appear, rather than one.
In the limit $z\rightarrow (1+ i\epsilon) w$, $X^L(z)\rightarrow X^L(w)$. Thus,  
the untwisted states in the l.h.s. of (\ref{OPEuutt}) combine to an intermediate state $e^{i(k_L+p_L)X^L}$, 
of left momentum $k_L+p_L$.
In the limit $z\rightarrow (1- i\epsilon) w$, $X^L(z)\rightarrow -X^L(w)$, and 
the untwisted states in the l.h.s. of (\ref{OPEuutt}) combine to an intermediate state $e^{i(k_L-p_L)X^L}$, 
of left momentum $k_L-p_L$.
A similar discussion holds
on the antiholomorphic side. Therefor,
 the untwisted states combine to form a state of left 
and right momentum $k_{L,R}+p_{L,R}$, and a state of left and right momentum $k_{L,R}-p_{L,R}$.
This is to be expected, since
every untwisted state of momentum $p_{L,R}$ includes also a piece carrying momentum 
$-p_{L,R}$, since its vertex operator is ${1\over 2}(e^{ip_{L,R}\cdot X^{L,R}}+e^{-ip_{L,R}\cdot X^{L,R}})$.

Fixing $w=1$ we get
\begin{eqnarray}
{\cal A}_{tt,p,k}&=& -\int dzd\bar{z} \,
\left(p_L\cJ k_L\right)\left(p_R\cJ k_R\right) \left|{z^{-1/2}\over z-1}\right|^2
\left({1-z^{1/2}\over 1+z^{1/2}}\right)^{2k_L\cdot p_L} 
\left({1-\bar{z}^{1/2}\over 1+\bar{z}^{1/2}}\right)^{2k_R\cdot p_R} \nn
&+& \left(p_{L,R} \leftrightarrow -p_{L,R} \right) + \left(k_{L,R} \leftrightarrow -k_{L,R} \right)
+ \left(p_{L,R},k_{L,R} \leftrightarrow -p_{L,R},-k_{L,R} \right) \nn
&=& -2 \int dzd\bar{z} \, \left(p_L\cJ k_L\right)\left(p_R\cJ k_R\right) |z(z-1)^2|^{-1} \nn
&\cdot&\left[
\left({1-z^{1/2}\over 1+z^{1/2}}\right)^{2k_L\cdot p_L} 
\left({1-\bar{z}^{1/2}\over 1+\bar{z}^{1/2}}\right)^{2k_R\cdot p_R} 
+ \left({1-z^{1/2}\over 1+z^{1/2}}\right)^{-2k_L\cdot p_L} 
\left({1-\bar{z}^{1/2}\over 1+\bar{z}^{1/2}}\right)^{-2k_R\cdot p_R}
\right] \ .
\nn
\end{eqnarray}

Denote 
\begin{equation}
{\cal A}_{tt,p,k}= -2\left(p_L\cJ k_L\right)\left(p_R\cJ k_R\right) 
\bigg[I\left(2k_L\cdot p_L,2k_R\cdot p_R\right) + 
I\left(-2k_L\cdot p_L,-2k_R\cdot p_R\right)\bigg] \ ,
\label{apAII}
\end{equation}
where
\begin{equation}
I(a_L, a_R)\equiv \int dzd\bar{z} \, |z(1-z)^2|^{-1} 
\left({1-z^{1/2}\over 1+z^{1/2}}\right)^{a_L} \left({1-\bar{z}^{1/2}\over 1+\bar{z}^{1/2}}\right)^{a_R} \ .
\end{equation}
Then,
substituting $z^\prime=z^{1/2}$ (with $z^\prime$ in the upper complex plane), $v={1+z^\prime\over 2}$,
we have
\begin{eqnarray}
I(a_L, a_R) &=& 4 \int_{{\rm Im} (z^\prime)\ge 0} dz^\prime d\bar{z}^\prime \, 
(1-z^\prime)^{a_L-1} (1+z^\prime)^{-a_L-1} 
(1-\bar{z}^\prime)^{a_R-1} (1+\bar{z}^\prime)^{-a_R-1} \nn
&=& \int_{{\rm Im} (v)\ge 0} dvd\bar{v} \, (1-v)^{a_L-1} v^{-a_L-1} (1-\bar{v})^{a_R-1} \bar{v}^{-a_R-1} \ .
\end{eqnarray}
Taking $a_{L,R} \rightarrow -a_{L,R}$ is equivalent to the exchange
 $v \leftrightarrow 1-v$. 
When $a_L-a_R$ is an integer we may write
\begin{eqnarray}
I(a_L, a_R) + I(-a_L, -a_R) &=& 
\int_{\bC} dvd\bar{v} \, (1-v)^{a_L-1} v^{-a_L-1} (1-\bar{v})^{a_R-1} \bar{v}^{-a_R-1} \nn
&=& \pi {\Gamma(-a_L)\Gamma(a_L)\Gamma(1) \over \Gamma(0) \Gamma(1-a_R) \Gamma(1+a_R)} = 0 \ .
\label{apIaLaR}
\end{eqnarray}

Since $k_L \cdot p_L - k_R \cdot p_R$ must be an integer for the theory to be modular invariant 
(as in any string theory with a toroidal compactification), we can  use (\ref{apIaLaR}) to calculate the 
scattering amplitude (\ref{apAII}). We thus get
\begin{eqnarray}
{\cal A}_{tt,p,k} &=& 0 \ .
\label{uutt}
\end{eqnarray}
All the above is
easily generalized to the non-compact background $\bC^2/\bZ_2$ by setting $p_L=p_R$, 
$k_L=k_R$.

\subsection{Study of $I(a_L,a_R)$}

We may perform the integration in $I(2k_L\cdot p_L, 2k_R\cdot p_R)$ 
as follows. Taking $u=1-v$ and denoting $a_{L,R} \equiv 2k_{L,R}\cdot p_{L,R} = (k_{L,R}+p_{L,R})^2$, then
for $a_L \ne a_R$,
\begin{eqnarray}
I(a_L,a_R) &=& \int_{{\rm Im}(u)\le 0} dud\bar{u} u^{a_L-1} (1-u)^{-a_L-1} \bar{u}^{a_R-1} (1-\bar{u})^{-a_R-1} 
= \int_{{\rm Im}(x)\le 0} dxd\bar{x} x^{a_L-1} \bar{x}^{a_R-1} \nn
&=& \int_\pi^{2\pi} d\phi  e^{i\phi (a_L-a_R)} \int_0^R dr r^{a_L+a_R-1}
= -i {1 - e^{i\pi(a_L-a_R)} \over a_L - a_R} \int_0^R dr r^{a_L+a_R-1} = 0 \ ,
\end{eqnarray}
where we have substituted $x={u\over 1-u}$, and later changed to polar coordinates. 
The result is zero because $a_L-a_R$ must be an even integer.
We thus see that $I$ has no contributions from $a_L \ne a_R$.

For $a_L=a_R \equiv a$
\begin{eqnarray}
I(a,a) &=& \int_{{\rm Im}(u)\le 0}  dud\bar{u} |u|^{2a-2} |1-u|^{-2a-2} = \int_{{\rm Im}(x)\le 0} dxd\bar{x} |x|^{2a-2} \nn
&=& \int_{|x|\le 1,{\rm Im}(x)\le 0} dxd\bar{x} |x|^{2a-2} + \int_{|x|\le 1,{\rm Im}(x)\le 0} dxd\bar{x} |x|^{-2a-2} \nn
&=& \pi \int_0^1 dr~r^{2a-1} + \pi \int_0^1 dr~r^{-2a-1} = \pi{r^{2a}\over 2a}\bigg|_0^1 + \pi{r^{-2a}\over -2a}\bigg|_0^1 
\label{Iaa}
\end{eqnarray}
where we have substituted $x={u\over 1-u}$ in the first line, $x\rightarrow -1/x$ in the second term of the second line, 
and $r=|x|$ in the last line.

The first term in the result of (\ref{Iaa}) is a pole at $a=0$ with the branch cut at $a<0$, coming from the limit 
$u\rightarrow 0$ (or $z\rightarrow 1+i\epsilon$). $a = (p_L+k_L)^2 = (p_R+k_R)^2$, so 
this corresponds to an exchange of an untwisted state of left and right momenta $p_L+k_L,p_R+k_R$ with 
$(p_L+k_L)^2 = (p_R+k_R)^2 =0$.
The second term in the result of (\ref{Iaa}) is a pole at $a=0$ with the branch cut at $a>0$, 
coming from the limit 
$u\rightarrow 1$ (or $z\rightarrow 1-i\epsilon$). $-a = (p_L-k_L)^2 = (p_R-k_R)^2$, so 
this corresponds to an exchange of an untwisted state of left and right momenta $p_L-k_L,p_R-k_R$ with 
$(p_L-k_L)^2 = (p_R-k_R)^2 =0$.
Identical poles arise from $I(-2k_L\cdot p_L, -2k_R\cdot p_R)$. Therefor only poles in 
$(p_{L,R}\pm k_{L,R})^2 = 0$ contribute to the scattering amplitude.

\section{Twist Fields in Open Strings}

Consider open strings on the upper half plane, ${\rm Im}z\ge 0$,
with the boundary conditions 
$\del X(z)=D \bar{\del}X(\bar{z})$ ($D=\pm1$) at $z=\bar{z}$. 
Close to the worldsheet boundary, (\ref{clBOPE}) 
must be corrected by a boundary term. We may write:
\begin{eqnarray}
\del X(z)\sigma(w,\bar{w}) \sim f_1(z,w,\bar{w})\tau (w,\bar{w})  &,&
\bar{\del} X(\bar{z})\sigma(w,\bar{w}) \sim f_2(\bar{z},w,\bar{w}) \bar{\tau} (w,\bar{w})  \ .
\end{eqnarray}
$f_{1,2}$ must be multiplied by $-1$ as $z$ ($\bar{z}$) rotates around $w$ ($\bar{w}$).
Since $\sigma(w,\bar{w})$ corresponds to the unexcited twisted state we must have 
\begin{eqnarray}
f_1(z,w,\bar{w}) = (z-w)^{-1/2}\, h_1(z,w,\bar{w})   &,&
f_2(\bar{z},w,\bar{w}) = (\bar{z}-\bar{w})^{-1/2}\, h_2(\bar{z},w,\bar{w})  \ ,
\end{eqnarray}
such that
\begin{equation}
h_{1}(z,w,\bar{w})|_{z\rightarrow w}=\mbox{finite} \ ,~~~
h_{2}(\bar{z},w,\bar{w})|_{\bar{z}\rightarrow \bar{w}}=\mbox{finite} \ .
\end{equation}
The boundary conditions now imply 
\begin{eqnarray}
h_1(z,w,\bar{w}) &=& (z-\bar{w})^{-1/2}\,\tilde{h}(w,\bar{w})  \ ,  \nn
h_2(\bar{z},w,\bar{w}) &=& i (\bar{z}-w)^{-1/2}\,\tilde{h}(w,\bar{w})  \ .
\end{eqnarray}
and $\bar{\tau} (w,\bar{w}) = -iD\tau (w,\bar{w})$.
One can take $\tilde{h}(w,\bar{w})=(w-\bar{w})^{+1/2}$ so that the OPE
reduces to that on $S^2$ as $z\rightarrow w$.
We then have
\begin{eqnarray}
\del X(z)\sigma(w,\bar{w}) &\sim& (z-w)^{-1/2}(z-\bar{w})^{-1/2}
(w-\bar{w})^{+1/2}\,\tau (w,\bar{w})  \ , \nn
\bar{\del} X(\bar{z})\sigma(w,\bar{w}) &\sim& D(\bar{z}-\bar{w})^{-1/2}
(\bar{z}-w)^{-1/2}(w-\bar{w})^{+1/2}\, \tau(w,\bar{w})  \ .
\end{eqnarray}

\section{Two Twist Fields Correlation Function on $D_2$}

For every complex dimension we may write
\begin{eqnarray}
-{1\over 2}{\langle \partial_z X(z) \partial_w \bar{X}(w)\sigma(z_1,\bar{z_1})\sigma(z_2,\bar{z_2}) \rangle_{D_2} \over 
\langle \sigma(z_1,\bar{z_1})\sigma(z_2,\bar{z_2}) \rangle_{D_2}} &=& \nn
(z-z_1)^{-1/2} (z-z_2)^{-1/2} (z-\bar{z}_1)^{-1/2} (z-\bar{z}_2)^{-1/2} &\times& \nn
(w-z_1)^{-1/2} (w-z_2)^{-1/2} (w-\bar{z}_1)^{-1/2} (w-\bar{z}_2)^{-1/2} 
&\times& g(z,w,z_i,\bar{z}_i) \ , 
\label{apXXTT}
\end{eqnarray}
where the function $g(z,w,z_i,\bar{z}_i)$ is single-valued and is holomorphic in $z$ and $w$.
As $z\rightarrow w$ the correlation function 
must have a double pole with residue one, and no single pole.
Thus, it should behave as $\sim 1/(z-w)^2$ plus finite terms.

(\ref{apXXTT}) must be invariant under worldsheet translations parallel to the worldsheet boundary
($(z,\bar{z}\rightarrow z+a,\bar{z}+a$ for real $a$).
After performing the coordinate transformations 
$z\rightarrow -1/z,~w\rightarrow -1/w$ and
$z\rightarrow \lambda z,~w\rightarrow \lambda w$
(for real $\lambda$) (\ref{apXXTT}) must be multiplied by $z^2w^2$ and
 $\lambda^{-2}$ respectively.
Thus
\begin{eqnarray}
g(z,w,z_i,\bar{z}_i) &=& {1\over 6(z-w)^2} [(z-z_1)(z-z_2)(w-\bar{z}_1)(w-\bar{z}_2)
+(z-z_1)(z-\bar{z}_1)(w-z_2)(w-\bar{z}_2) \nn
&+&(z-z_1)(z-\bar{z}_2)(w-\bar{z}_1)(w-z_2)+(z-\bar{z}_1)(z-z_2)(w-z_1)(w-\bar{z}_2) \nn
&+&(z-\bar{z}_1)(z-\bar{z}_2)(w-z_1)(w-z_2)+(z-z_2)(z-\bar{z}_2)(w-z_1)(w-\bar{z}_1)] \nn
&+&A(z_1-z_2)(\bar{z}_1-\bar{z}_2)+B(z_1-\bar{z}_2)(\bar{z}_1-z_2) \ ,
\label{XXTThol}
\end{eqnarray}
where $A$ and $B$ are constants.
Subtracting $1/(z-w)^2$ and taking the limit $z\rightarrow w$ we get
\begin{eqnarray}
&&{\langle T(z)\sigma(z_1,\bar{z_1})\sigma(z_2,\bar{z_2}) \rangle \over 
\langle \sigma(z_1,\bar{z_1})\sigma(z_2,\bar{z_2}) \rangle} = \nn
&-&{1\over 12} [(z-\bar{z}_1)^{-1}(z-\bar{z}_2)^{-1}
+(z-\bar{z}_1)^{-1}(z-z_2)^{-1}+(z-z_2)^{-1}(z-\bar{z}_2)^{-1} \nn
&+&(z-z_1)^{-1}(z-\bar{z}_1)^{-1}+(z-z_1)^{-1}(z-\bar{z}_2)^{-1}
+(z-z_1)^{-1}(z-z_2)^{-1}] \nn
&+&{1\over 8}\left[(z-z_1)^{-2}+(z-z_2)^{-2}+(z-\bar{z}_1)^{-2}+(z-\bar{z}_2)^{-2}\right] \nn
&+&(z-z_1)^{-1}(z-z_2)^{-1}(z-\bar{z}_1)^{-1}(z-\bar{z}_2)^{-1} 
\left(A(z_1-z_2)(\bar{z}_1-\bar{z}_2)+B(z_1-\bar{z}_2)(\bar{z}_1-z_2)\right) \ . \nn
\label{enTT}
\end{eqnarray}
In the limit $z\rightarrow z_1$, the coefficient of $(z-z_1)^{-2}$ is the twist field conformal dimension $1/8$, 
and the coefficient of the $(z-z_1)^{-1}$ term is 
$\partial_{z_1} \ln \langle \sigma(z_1,\bar{z_1})\sigma(z_2,\bar{z_2})\rangle_{D_2}$.
Therefor,
\begin{eqnarray}
\partial_{z_1} \ln \langle \sigma(z_1,\bar{z_1})\sigma(z_2,\bar{z_2})\rangle_{D_2}
&=& -{1\over 12}\left((z_1-z_2)^{-1}+(z_1-\bar{z}_1)^{-1}+(z_1-\bar{z}_2)^{-1}\right)\nn
&+& A\left({1\over z_1-\bar{z}_1}-{1\over z_1-\bar{z}_2}\right)
+ B\left({1\over z_1-\bar{z}_1}-{1\over z_1-z_2}\right) \ .
\end{eqnarray}
Repeating this procedure for the antiholomorphic part of the energy-momentum tensor and integrating, we get
\begin{equation}
\langle \sigma(z_1,\bar{z_1})\sigma(z_2,\bar{z_2})\rangle_{D_2} = 
(z_1-\bar{z_1})^{A+B-{1\over 12}}(z_1-z_2)^{-B-{1\over 12}}
(z_1-\bar{z_2})^{-A-{1\over 12}}\, \times\, ({\mbox c.c.}) \ .
\end{equation}
We have used the fact that the holomorphic and antiholomorphic parts must have the same constants $A$, $B$ 
so that $T(z)=\bar{T}(\bar{z})$ on the worldsheet boundary.


\begin{thebibliography}{99}

\bibitem{ademollo;76}
M.~Ademollo {\it et al.},
``Dual String With U(1) Color Symmetry,''
Nucl.\ Phys.\ B {\bf 111}, 77 (1976).

\bibitem{Gates:1988tn}
S.~J.~Gates, L.~Lu and R.~N.~Oerter,
``Simplified SU(2) Spinning String Superspace Supergravity,''
Phys.\ Lett.\ B {\bf 218}, 33 (1989).


\bibitem{Ooguri:1991fp}
H.~Ooguri and C.~Vafa,
``Geometry of N=2 strings,''
Nucl.\ Phys.\ B {\bf 361}, 469 (1991).


\bibitem{Cheung:2002yw}
Y.~K.~Cheung, Y.~Oz and Z.~Yin,
``Families of N = 2 strings,''
JHEP {\bf 0311}, 026 (2003)
[arXiv:hep-th/0211147].


\bibitem{Gluck:2003wg}
D.~Gluck, Y.~Oz and T.~Sakai,
``The effective action and geometry of closed N = 2 strings,''
JHEP {\bf 0307}, 007 (2003)
[arXiv:hep-th/0304103].


\bibitem{Hull:1996zt}
C.~M.~Hull,
``The geometry of N = 2 strings with torsion,''
Phys.\ Lett.\ B {\bf 387}, 497 (1996)
[arXiv:hep-th/9606190].

\bibitem{Gluck:2003pa}
D.~Gluck, Y.~Oz and T.~Sakai,
``D-branes in N = 2 strings,''
JHEP {\bf 0308}, 055 (2003)
[arXiv:hep-th/0306112].

\bibitem{Ohta:1989pr}
N.~Ohta and S.~Osabe,
``Hidden Extended Superconformal Symmetries in Superstrings,''
Phys.\ Rev.\ D {\bf 39}, 1641 (1989).

\bibitem{Berkovits:1994vy}
N.~Berkovits and C.~Vafa,
``N=4 topological strings,''
Nucl.\ Phys.\ B {\bf 433}, 123 (1995)
[arXiv:hep-th/9407190].


\bibitem{Aharony:2003vk}
O.~Aharony, B.~Fiol, D.~Kutasov and D.~A.~Sahakyan,
``Little string theory and heterotic/type II duality,''
Nucl.\ Phys.\ B {\bf 679}, 3 (2004)
[arXiv:hep-th/0310197].

\bibitem{bl:1996}
J.~Bischoff and O.~Lechtenfeld,
``Path-Integral Quantization of the (2,2) String,''
Int.\ J.\ Mod.\ Phys.\ A {\bf 12}, 4933 (1997)
[arXiv:hep-th/9612218].


\bibitem{Mathur:1986cz}
S.~D.~Mathur and S.~Mukhi,
``Brst Quantization Of Twisted Extended Fermionic Strings,''
Phys.\ Rev.\ D {\bf 36}, 465 (1987).


\bibitem{klp}
S.V.~Ketov, O.~Lechtenfeld and A.J.~Parkes,
``Twisting the N=2 Strings,''
Phys.\ Rev.\ D {\bf 51}, 2872 (1995)
[arXiv:hep-th/9312150].


\bibitem{Friedan:1985ge}
D.~Friedan, E.~J.~Martinec and S.~H.~Shenker,
``Conformal Invariance, Supersymmetry And String Theory,''
Nucl.\ Phys.\ B {\bf 271}, 93 (1986).


\bibitem{JL:1997}
K.~Junemann and O.~Lechtenfeld
``Chiral BRST Cohomology of N=2 Strings at Arbitrary Ghost and Picture Number,''
Commun.\ Math.\ Phys. {\bf 203}, 53 (1999)
[arXiv:hep-th/9712182].


\bibitem{Bienkowska:1991zs}
J.~Bienkowska,
``The Generalized no ghost theorem for N=2 SUSY critical strings,''
Phys.\ Lett.\ B {\bf 281}, 59 (1992)
[arXiv:hep-th/9111047].


\bibitem{Dixon:1987}
L.~J.~Dixon, D.~Friedan, E.~J.~Martinec and S.~H.~Shenker,
``The Conformal Field Theory Of Orbifolds,''
Nucl.\ Phys.\ B {\bf 282}, 13 (1987).


\bibitem{JS:01}
K.~Junemann and B.~Spendig,
``D-brane Scattering of N=2 Strings,''
Phys.\ Lett.\ B {\bf 520}, 163 (2001)
[arXiv:hep-th/0108069].


\bibitem{mar;92}
N.~Marcus,
``The N=2 open string,''
Nucl.\ Phys.\ B {\bf 387}, 263 (1992)
[arXiv:hep-th/9207024].


\bibitem{Douglas:1996sw}
M.~R.~Douglas and G.~Moore,
``D-branes, Quivers, and ALE Instantons,''
arXiv:hep-th/9603167.


\bibitem{allloop}
H.~Ooguri and C.~Vafa,
``All Loop N=2 String,''
Nucl.\ Phys.\ B {\bf 451}, 121 (1995)
[arXiv:hep-th/9505183].


\bibitem{FT:1981}
E.S.~Fradkin and A.A.~Tseytlin,
``Quantization of Two-Dimensional Supergravity
 and Critical Dimensions for String Models,''
Phys. Lett. {\bf 106B}, 063 (1981).


\bibitem{Li:1992rr}
M.~Li,
``Gauge symmetries and amplitudes in N=2 strings,''
Nucl.\ Phys.\ B {\bf 395}, 129 (1993)
[arXiv:hep-th/9204027].


\bibitem{Schwimmer:1986mf}
A.~Schwimmer and N.~Seiberg,
``Comments On The N=2, N=3, N=4 Superconformal Algebras In Two-Dimensions,''
Phys.\ Lett.\ B {\bf 184}, 191 (1987).




\end{thebibliography}
\end{document}